\documentstyle{article}

\begin{document}
\baselineskip = 20pt
\input epsf

\ifx\epsfbox\UnDeFiNeD\message{(NO epsf.tex, FIGURES WILL BE IGNORED)}
\def\figin#1{\vskip2in}
\else\message{(FIGURES WILL BE INCLUDED)}\def\figin#1{#1}\fi

\def\ifig#1#2#3{\xdef#1{fig.~\the\figno}
\vskip 0.3cm \goodbreak\figin{\centerline{#3}}%
\vskip 0.5cm\smallskip\centerline{\vbox{\baselineskip12pt
\advance\hsize by -1truein\noindent{\bf
Fig.~\the\figno:} #2}}
\bigskip\global\advance\figno by1}

\def\figures{\centerline{{\bf Figure
Captions}}\medskip\parindent=40pt%
\def\fig##1##2{\medskip\item{FIG.~##1.  }##2}}
\newwrite\ffile\global\newcount\figno \global\figno=1
\def\fig{fig.~\the\figno\nfig}
\def\nfig#1{\xdef#1{fig.~\the\figno}%
\writedef{#1\leftbracket fig.\noexpand~\the\figno}%
\ifnum\figno=1\immediate\openout\ffile=figs.tmp\fi\chardef\wfile=
\ffile%
\immediate\write\ffile{\noexpand\medskip\noexpand\item{Fig.\
\the\figno. }
\reflabeL{#1\hskip.55in}\pctsign}\global\advance\figno by1\findarg}

\parindent 25pt
\overfullrule=0pt
\tolerance=10000
\def\Re{\rm Re}
\def\Im{\rm Im}
\def\titlestyle#1{\par\begingroup \interlinepenalty=9999
     \fourteenpoint
   \noindent #1\par\endgroup }
\def\tr{{\rm tr}}
\def\Tr{{\rm Tr}}
\def\half{{\textstyle {1 \over 2}}}
\def\quart{{\textstyle {1 \over 4}}}
\def\calt{{\cal T}}
\def\ie{{\it i.e.}}
\def\np{Nucl. Phys.}
\def\pl{Phys. Lett.}
\def\pr{Phys. Rev.}
\def\prl{Phys. Rev. Lett.}
\def\cmp{Comm. Math. Phys.}
\def\quart{{\textstyle {1 \over 4}}}
\def\RR{${\rm R}\otimes{\rm R}~$}
\def\NSNS{${\rm NS}\otimes{\rm NS}~$}
\def\RNS{${\rm R}\otimes{\rm NS}~$}
\def\calI{{\cal I}}
\def\be{\begin{equation}}
\def\ee{\end{equation}}
\def\spin32{{\rm Spin}(32)/Z_2}

\baselineskip=14pt

\newskip\humongous \humongous=0pt plus 1000pt minus 1000pt
\def\caja{\mathsurround=0pt}
\def\eqalign#1{\,\vcenter{\openup1\jot \caja
	\ialign{\strut \hfil$\displaystyle{##}$&$
	\displaystyle{{}##}$\hfil\crcr#1\crcr}}\,}
\newif\ifdtup
\def\panorama{\global\dtuptrue \openup1\jot \caja
	\everycr{\noalign{\ifdtup \global\dtupfalse
	\vskip-\lineskiplimit \vskip\normallineskiplimit
	\else \penalty\interdisplaylinepenalty \fi}}}
\def\eqalignno#1{\panorama \tabskip=\humongous
	\halign to\displaywidth{\hfil$\displaystyle{##}$
	\tabskip=0pt&$\displaystyle{{}##}$\hfil
	\tabskip=\humongous&\llap{$##$}\tabskip=0pt
	\crcr#1\crcr}}
\relax

\thispagestyle{empty}

\null
\vskip-2.0cm
{\hfill  CPTH/S570.1297}

{\hfill DAMTP/97-50}

{\hfill WIS/97-34}

\vskip 0.1cm
{\hfill hep-th/9712086}
\vskip 1.5 cm

\begin{center}
{\large\bf  (8,0) Quantum mechanics and symmetry  }
\centerline{\large\bf  enhancement in 
 type I'  superstrings }
\break

 \centerline{ Constantin P. Bachas$\ ^a$,  Michael
B. Green$\ ^b$ and Adam  Schwimmer$\ ^c$}

\vskip  0.5in
\centerline{$\ ^a$\  Centre de Physique Th\'eorique,  Ecole
Polytechnique,}
\centerline{ 91128 Palaiseau, France}
\centerline{bachas@pth.polytechnique.fr}
 \vskip 0.3cm

\centerline{$\ ^b$ Department of Applied Mathematics and
Theoretical Physics,}
\centerline{Silver Street,Cambridge CB3 9EW, United Kingdom}
\centerline{M.B.Green@damtp.cam.ac.uk}
\vskip 0.3cm

\centerline{$\ ^c$ Department of Physics of Complex Systems, }
\centerline{Weizmann Institute of Science, Rehovot, 76100, Israel}
\centerline{ftschwim@wicc.weizmann.ac.il}

\end{center}
\vskip 1.0 cm
\rm

\begin{quote}
{\bf ABSTRACT:}

The low-energy supersymmetric quantum mechanics
describing D-particles in the background of D8-branes
 and orientifold planes
is analyzed in detail, including a careful discussion of
Gauss'  law and normal ordering of operators.
This  elucidates the mechanism that binds
D-particles to an  orientifold plane,
in accordance with the predictions of
heterotic/type I duality.   The ocurrence of enhanced symmetries
 associated with massless  bound states of a D-particle with one
  orientifold plane  is  illustrated by the enhancement of
 $SO(14) \times U(1)$ to $E_8$ and $SO(12)\times U(1)$ to $E_7$ at
strong  type I' coupling.  
Enhancement to higher-rank groups involves both orientifold planes.
  For example, the enhanced  $E_8 \times E_8 \times SU(2)$ symmetry at
 the self-dual radius of the heterotic string is  seen as the result
 of two D8-branes coinciding midway between the orientifold planes, while
the enhanced  $SU(18)$ symmetry results from the coincidence of all sixteen
D8-branes and  $SO(34)$ when  they  also  coincide with an orientifold plane.   
   As a separate by-product,  the s-rule 
of brane-engineered gauge theories
  is  derived by relating it through a chain of
dualities to  the  Pauli exclusion principle.

\end{quote}
\vskip1cm

\normalsize

\newpage
\pagestyle{plain}
\setcounter{page}{1}

\def\baselinestretch{1.2}
\baselineskip 16 pt
\noindent
 \setcounter{equation}{0}

\section{ Introduction}

  The dynamics of D-particles in the presence of D8-branes and
orientifold planes plays  an important part  in type I'
superstring theory
\cite{polwitt}, which is the compactification of M-theory  on a cylinder
\cite{horwitt}. Several  features of the associated (8,0) supersymmetric
quantum
mechanics \footnote{The notation $(8,0)$ defines the number of  
left-moving
 and right-moving supersymmetries in the
 T-dual $(1+1)$-dimensional field theory
 describing strings in the presence of D9-branes.}
 were  introduced in \cite{DanFer,banksetal} and further
discussed
in the context of heterotic/type-I duality
and of the matrix conjecture in
\cite{KS,Lowe,Banmot,rey,hor,KR,mall,last}.
In this   paper several aspects of the dynamics of this system will be
 considered in more detail.    Particular care will be taken  with issues
 relating to normal ordering, the Gauss law constraints on physical
states
 and the effect of  the anomalous creation  of fundamental strings
 \cite{bdg,klebanov,bgl,alwis,howu,ohta}
when a D-particle crosses a  D8-brane.  Several of our conclusions
  differ from those in  the  previous papers on this subject.
  We  will be  lead to a rather subtle pattern of level crossing
 as the moduli of the system are varied. This will elucidate the
  mechanism responsible for
binding D-particles to orientifold planes,  as  required by
heterotic-type I duality.
 The expected   enhanced symmetries of the type I' theory  \cite{polwitt}
at strong coupling  arise  explicitly  at points where certain
bound-state masses vanish.

In section 2 we will review the quantum  mechanical formulation of  the
dynamics of D-particles in the type IIA superstring theory   in the  
presence of
a  D8-brane.  The quantum-mechanical approximation to the full   
dynamics
should be   good   for slowly-moving D-particles and is motivated  
by explicit
string calculations.   A D8-brane  acts like a domain wall and  
separates
regions of space in which the mass parameter, $m$, of the IIA  
theory differs by
one unit.  We will see that the value of $m$  is correlated, via  
Gauss' law,
with the number of strings joining the D-particle to the  
D8-branes, which  is in accord with the picture that a string is
created or destroyed when a D-particle passes through a  
D8-brane. The fact that these  strings have a unique BPS ground state which
 is  fermionic 
leads, via a chain of dualities,
 to an understanding of the s-rule of Hanany and Witten
\cite{hw,egk,vo,egkrs}. This rule  states that a configuration with  more
 than one D3-brane joining a NS-fivebrane   and a D5-brane, and extending
in all space-time dimensions, 
cannot have a quantum state  with unbroken supersymmetry.

Our  considerations are extended in Section 3 to  type I'  
D-particle  quantum
mechanics, which is a projection of the type IIA theory.  This is  again
motivated by string calculations,  now   in nine dimensions and in  the
presence of two orientifold planes  and  sixteen parallel  
D8-branes with
their mirror images.   The presence of states of non-zero winding  
restrict the
region of moduli space in which it is consistent to use the quantum  
mechanical
truncation of this system.  This is the region in which one of the  
orientifold
planes is at the origin   in the transverse direction with   half the
D8-branes  close by while the other orientifold and the remaining
D8-branes are taken to infinity.  A  single D-particle in this  
background is
necessarily stuck to the orientifold plane and  generically  has sub-threshold  
BPS bound
states with masses that are easy to determine.   We will also discuss the
states of  a  probe D-particle and its mirror as they move in the  
transverse
directions.    Gauss' law again leads to
 strong constraints on the  pattern of strings joining the  
D-particle to the
various branes  and again there are  explicit sub-threshold BPS    
bound states
of the D-particle pair with the orientifold plane with easily calculable  
masses.

  In section 4 we will see that the masses of these  type I' BPS    
sub-threshold
bound states have  precisely  the values  expected on the grounds of
heterotic/type I duality.  We will first discuss symmetry enhancement
involving only one orientifold and its nearby eight D8-branes, which
can be studied consistently within the quantum mechanical approximation
to the system. We will find  agreement with the predictions of duality. 
The description of symmetry enhamcement in  the region of moduli space in  
which the two orientifold planes are at finite separations  typically requires the inclusion of string winding modes and falls   
outside the range of validity of the simple quantum mechanical approximation to type I' dynamics.  Nevertheless, it is straightforward to determine the physical ground states in this region by continuing from the region in which quantum mechanics is valid for each orientifold independently.    
An example of such an enhancement is the 
point at which the symmetry becomes $E_8 \times E_8 \times SU(2) \times U(1)$.
  This is the 
symmetry of the nine-dimensional heterotic string compactified to its 
  self-dual radius in the absence of 
any Wilson lines.  We will find this symmetry arising in the type I'
 picture when seven D8-branes (and 
their mirrors) coincide with each orientifold plane and the remaining
 two D8-branes coincide at a 
point  in between the orientifold planes (with similar coincidence of
 their mirrors). The
 symmetry enhancement
 to $SU(18) \times U(1)$  can be likewise seen to arise 
 when all sixteen D8-branes coincide half-way between the 
two orientifold planes.
Another special example is the point at which 
all D8-branes coincide with one
 orientifold plane and the symmetry is enhanced to $SO(34) \times U(1)$ at
 a particular value of the coupling.

\section{D-particles and  D8-branes}

In this section we will  reconsider the   dynamics of D-particles
interacting
with  a  D8-brane.  Although many of our considerations only
depend on the
low-energy quantum-mechanical degrees of freedom arising from the string
ground-state degrees of freedom,  it is essential for some
purposes to bear in
mind that the complete string theory also  contains winding states and
oscillator excitations.

\subsection{(8,8) Quantum mechanics}

 Consider first  a cloud of $n$  isolated
  type-IIA D-particles moving slowly in the ten-dimensional IIA vacuum.
 They are  described at sub-stringy energies
by the usual  supersymmetric $U(n)$ quantum mechanics
\cite{Halp,deW,matrix}
 with lagrangian
\begin{equation}
{\cal L}  = {1\over 2} \; \tr\; \Bigl( D Y_j DY_j + {g^2\over 2}
[ Y_i, Y_j]^2
+ i \;
  S_A DS_A +  g\; S_A (\Gamma^0\Gamma^j)_{AB} [ Y_j, S_B]
 \Bigr) \ ,
\label{lag}
\end{equation}
which is invariant under the (8,8) supersymmetry transformations,
\begin{equation}
\eqalign{
&\delta Y^j = i \xi_A (\Gamma^0 \Gamma^j)_{AB} S_B \cr
&\delta A_0 = -i \xi_A  S_A \cr
&\delta S_A =  DY_j (\Gamma^0\Gamma^j)_{AB} \xi_B
+ {ig\over 4} [Y_i, Y_j ]  [\Gamma^i,\Gamma^j]_{AB}
 \xi_B. \cr}
\label{transfo}
\end{equation}
The covariant derivative of any of the hermitian  $n\times n$  
matrices,  $Y_j$,
$S_A$ and $A_0$, is defined by
$D M  = {\dot M} + i g [A_0, M]$.
Under the $SO(1,9)$ Lorentz group $i,j=1,..,9$ are spatial-vector
indices,  and  $A,B= 1,..,16$ are Weyl-Majorana spinor indices.
 The ten-dimensional $\Gamma$-matrices are
purely imaginary and satisfy the algebra $\{\Gamma^\mu, \Gamma^\nu\}
= -2 \eta^{\mu\nu}$ with metric signature $(-+ ...+)$.   The   matrix
coordinates appearing in the lagrangian and the detailed form  of  
the above
expressions  originate from the supersymmetric Yang--Mills theory that
describes the ground states of the $(00)$ strings   joining pairs of
D-particles.

  Canonical quantization in the $A_0=0$ gauge
leads to the commutation relations
\begin{equation}
\eqalign{
&[ \Pi_i^{rs}, Y_j^{tq} ] = -i\; \delta_{ij} \delta^{rq}\delta^{st}
\cr
& \{ S_A^{rs}, S_B^{tq} \} =  \delta_{AB} \delta^{rq}\delta^{st} \ , \cr}
\end{equation}
where $\Pi_j = {\dot Y}_j$. The supercharges and the  hamiltonian
read
\begin{equation}
\eqalign{
& Q_A = (\Gamma^0\Gamma^j)_{AB}\;  {\rm tr} ( \Pi_j S_B )
+  {ig\over 4}  [\Gamma^i, \Gamma^j]_{AB} \;
 {\rm tr} ( [Y_i, Y_j] S_B)
\cr &
{\cal H}  = {1\over 2} \; \tr\; \Bigl( \Pi_j^2 -  {g^2\over 2}
[ Y_i, Y_j]^2 -  g\; S_A (\Gamma^0\Gamma^j)_{AB} [ Y_j, S_B] \Bigr),
\label{Ham}
 \cr}
\end{equation}
while Gauss' law is the matrix equation
\begin{equation}
{\cal G}= - {\delta{\cal L}\over \delta (g A_0)}
 = i\; [ \Pi_j, Y_j] -
 S_A  S_A - n \; {\bf 1}  = 0 \ .
\label{G}
\end{equation}
The supersymmetry algebra that follows from (\ref{Ham}) is given by
\begin{equation}
\{ Q_A, Q_B \} = 2\delta_{AB} {\cal H} -    2g (\Gamma^0\Gamma^i)_{AB}\;
{\rm tr}( Y_i {\cal G})\ ,
\label{sup}
\end{equation}
and  reduces to  standard  form  in the subspace of
physical states,  $<{\rm phys}^\prime\vert {\cal G} \vert {\rm
phys} > = 0$.

  An important fact about these expressions is that they are free
from   operator-ordering ambiguities. This is manifest for the  
supercharges
$Q_A$.  The potential ambiguity in the hamiltonian
 would have to be linear in  
$Y^i$ and is
therefore forbidden   by
Lorentz invariance. Closure of the algebra then   uniquely  fixes  the
ordering in Gauss' law to be the one of equation (\ref{G}).
The piece  proportional to the identity in this expression arises
because there are nine  bosonic  and sixteen  fermionic terms
that need to be rearranged in ${\delta{\cal L}/ \delta A_0}$.
The addition of a   Chern--Simons term,
$\int {\rm tr} A_0$,    is not allowed  by
 the supersymmetries of the problem.    In any case its  presence  
would  lead
to another inconsistency. The canonical
commutation relations imply  the operator identity
 \begin{equation}
{\rm tr}\;{\cal G} = 0\ ,
\end{equation}
which is consistent with the fact that there are no states charged under
the abelian $U(1)$ factor of $U(n)$.    The addition of a Chern--Simons  term in the action  would  lead to   an additional term
proportional to the identity  in this equation  which  would then have no solutions.
As a result  there are  no ambiguous operator-ordering parameters in this system
of isolated D-particles.

\subsection{Adding a  D8-brane}

Let us now  introduce  a  D8-brane transverse to the ninth direction. This breaks the rotational symmetry from $SO(9)$
to $SO(8)$. Henceforth we will reserve the indices $i,j$ for
$SO(8)$ vectors only. A  ten-dimensional Weyl-Majorana spinor
decomposes  as $8_s\oplus 8_c$. In this basis
$$
 \Gamma^1...\Gamma^8 =\pm \Gamma^0\Gamma^9 = \left(
\matrix{1&0\cr 0&-1\cr} \right) \ ,
$$
where the ten-dimensional chirality is + for $S_A$ and $\xi_A$, and  
$-$  for
$Q_A$ (in order that  $\bar\xi Q$ be  non-zero).
More explicitly, the matrix supercoordinates
of the cloud of D-particles are
$$
( Y_i, Y_9 ) \ \ \ {\rm and} \ \ \ (S_a , S_{\dot a} ) \ ,
$$
where $S_a$ ($S_{\dot a}$) has positive (negative) chirality under the
broken   $SO(1,1)$  that mixes the ninth coordinate and   time.
\footnote{The $SO(1,1)$  symmetry can be formally restored
by compactifying the ninth coordinate  on a circle of
vanishing radius. Under a   T-duality this configuration is
equivalent to
a system  of  n  parallel,  infinite D-strings  inside a nine-brane.
This  background is however inconsistent, because the nine-brane flux
has nowhere to escape.}
The supersymmetry algebra (\ref{sup})
 can be similarly decomposed as,
\begin{equation}
\eqalign{
& \{ Q_a, Q_b \} = 2\delta_{ab}\; \bigl(\;  {\cal H} + \;
 g\; {\rm tr}(Y_9{\cal G})\; \bigr) \cr &
\{ Q_{\dot a}, Q_{\dot b} \} = 2\delta_{{\dot a}{\dot b}}\;
 \bigl(\;  {\cal H} -   \;
 g\; {\rm tr}(Y_9{\cal G})\; \bigr) \cr&
\{ Q_a, Q_{\dot b} \} =
 2 g\; \gamma^i_{a \dot b}\; {\rm tr}(Y_i{\cal G})\; , \cr}
\end{equation}
where the $\gamma^i_{a \dot b}$ are the real matrices defined
in    \cite{GSW}.

Consider next  the (08) fundamental strings
which stretch between the D8-brane and the D-particles.
Such  strings  have  eight Neumann-Dirichlet
coordinates $x^i$,
one DD coordinate  $x^9$, and one NN coordinate  $x^0$.
The super Virasoro constraints
can be used to eliminate the  oscillator excitations of $x^0,x^9$
and of their fermionic world-sheet partners.
In the Ramond sector the  states are  built  from
half-integer moded oscillators
$x^{i}_{n+{1\over 2}}$ and $\psi^i_{n+{1\over 2}}$,
acting on a vacuum which is a spinor of $SO(1,1)$, but a singlet of
the  $SO(8)$ rotation group. When the D-particle  coincides with  the
D8-brane,
the mass of this Ramond ground state is zero.
 The GSO projection fixes  the product
of the $SO(1,1)$ chirality times the world-sheet
fermion parity.  We will choose a convention in which this sign is   negative.
In the  Neveu-Schwarz sector  physical states are  built
with  oscillators $x^i_{n+{1\over 2}}$ and $\psi^i_n$
($n> 0$)  acting on a ground state  of $ ({\rm Mass})^2 =
1/2\alpha^\prime$. The Neveu-Schwarz ground state is  a spinor of
$SO(8)$ but a
scalar of $SO(1,1)$. The GSO projection now fixes the product of the
world-sheet fermion parity and of the chirality of the $SO(8)$ spinor.
The first few low-lying states are listed in table 1.

\vskip 0.3cm

\begin{table}[htp]
\begin{center}
\begin{tabular}{|c|c||c|c|c|}
\hline
 & & & & \\
$({\rm Mass})^2 \alpha^\prime$
 & $SO(8)$\  {\rm rep.}& boson & +\ fermion & $-$\ fermion
\\
 & & & & \\  \hline
 & & & & \\
0 & {\bf 1} & & & $\vert \emptyset >_R$  \\
 & & & & \\ \hline
 & & & & \\
1/2 & {\bf 8} & $\vert\emptyset  >_{NS}$ &
$\psi^i_{1\over 2}\vert \emptyset >_R $ &
$ x^i_{1\over 2}\vert \emptyset >_R$
 \\
 & & & & \\  \hline
 & & & & \\
1 & ${\bf 8 \otimes 8}$&
$x^i_{1\over 2}\vert \emptyset >_{NS}$&
$\psi^i_{1\over 2} x^j_{1\over 2}\vert \emptyset  >_R$ &
$ x^i_{1\over 2} x^j_{1\over 2}\vert \emptyset >_R$ ;\
 \ $\psi^i_{1\over 2} \psi^j_{1\over 2}\vert\emptyset   >_R$
\\
& & & & \\  \hline
\end{tabular}
\end{center}
\caption{Spectrum of (08) strings. The mass does not include the
contribution from the   length of the stretched strings.
The states are classified as
bosons, + fermions and $-$  fermions, according to their transformation
properties under $SO(1,1)$. The GSO projection forces the product
of world-sheet fermion parity times the $SO(1,1)$ and $SO(8)$ spinor
chiralities to be negative. We have not specified in this table
which of the three possible
eight-dimensional representations of $SO(8)$
correspond  to each state, since supersymmetry transformations
do not commute with triality.
The strings are oriented  so all states are complex.}
\end{table}

  When the D-particles are close to the
D8-brane the  light (08) strings are described by  a  negative
chirality fermion  ($\chi_r$) in the fundamental representation of
$U(n)$.
The new effective  lagrangian,
\begin{equation}
{\cal L}^\prime = {\cal L}\;  +\;  {i} \chi^\dagger D \chi -
{g} \chi^\dagger Y^9 \chi\ - {gm}\; {\rm tr}(A_0+Y^9)\ ,
\label{lagr}
\end{equation}
 is invariant under half of the sixteen supersymmetries (\ref{transfo}),
namely those corresponding to  positive-chirality parameters  $\xi_a$.
The fermions $\chi$ do not transform.
This is characteristic of  chiral  supersymmetry  in two dimensions.
The hamiltonian,
supercharges and Gauss  constraint   read
\begin{eqnarray} \label{newham}
 {\cal H}^\prime & =& {\cal H} + {g} \chi^\dagger Y^9 \chi\ +
{gm}\;  {\rm tr}Y^9\\
Q_{\dot a}^\prime &=& Q_{\dot a}\label{newsusyy}  \\
{\cal G}^\prime_{rs} & =& {\cal G}_{rs} +  \chi^\dagger_s \chi_r
+{m}\; \delta_{rs} \ . \label{gausscona}
\end{eqnarray}
They obey  the supersymmetry algebra
\begin{equation}\label{newsusy}
\{ Q^\prime_{\dot a}, Q^\prime_{\dot b} \}
 = 2\delta_{{\dot a{\dot b}}}\; \bigl(\;  {\cal H}^\prime - \;
 g\; {\rm tr}(Y_9{\cal G}^\prime)\; \bigr).
\end{equation}
Although the $\chi$ fermions
 do not enter in the expressions for the
 supercharges,
the algebra closes because   their hamiltonian is cancelled by
the Gauss-constraint term on the right-hand-side \cite{banksetal}.
All other states in  table 1 are in
long multiplets of the supersymmetry algebra. The first excited level,
for example,  can be easily seen to form an irreducible representation
with the same content as a (8,8) vector multiplet. It  is therefore
a long (8,0) multiplet.
 The only BPS states of the (08) string are
those corresponding to the fermion $\chi$.   For most of the  
considerations in
this paper we will be concerned with the low-energy limit in which  
it will be
sufficient to consider the effective quantum mechanics of these  
ground states.

Although we are not presenting the details here, it is very easy to  
motivate the  quantum mechanical hamiltonian, (\ref{newham}), from  
an exact
string calculation in which the D-particle potential is determined by a
cylindrical world-sheet
with  one boundary  on the D-particle world-volume  and the other
on the
D8-brane.  This may be viewed as the partition function for open
strings
with one end on each brane. As with other BPS systems, this is  
equivalent to a
one-loop  open-string calculation in which only the BPS
ground-state fermions, $\chi_r$, contribute so the  expression coincides
precisely with
the energy in the quantum mechanical problem described above.
The standard stringy calculation of the phase shift \cite{b} shows
on the other hand that the $v^2$ force obtains  contributions from excited
states \cite{DanFer,bgl},  consistent with the fact that
 supersymmetry alone does not
determine the metric in the $Y^9$ direction.

A further important observation is that if the   D-particles are
 changed to anti-D-particles
in the above discussion   the sign of the GSO projection changes and
 hence  the sign of the
chirality of the fermions $\chi$ also changes.    However, the
  configuration is   still
supersymmetric under  the (0,8) supersymmetries corresponding to
 negative-chirality
parameters $\xi_{\dot a}$.  This should be contrasted
with other supersymmetric combinations   of Dp-branes and  
Dp'-branes,  where
changing the Dp-branes
to  anti Dp-branes breaks all supersymmetries \cite{J2}.   Of course, if
D-particles and anti-D-particles  are  simultaneously present in  
the D8-brane
background  all sixteen
supersymmetries are  broken.


\subsection{Gauss' law, string creation and the s-rule}

One  novel feature of the above  system is the appearance of the  
arbitrary
parameter $m$  and a corresponding
Chern--Simons term which  is  compatible with (8,0) supersymmetry.
Closure of the algebra   requires the same ordering for the  
$\chi$'s in the
hamiltonian
and in Gauss' law.   Since a  common reordering can be absorbed by  
shifting
the value of $m$
 there is no independent operator-ordering
  ambiguity in this system.\footnote{Supersymmetry  also allows 
 an arbitrary $Y^9$-dependent kinetic term in the direction transverse
to the brane \cite{DanFer,banksetal}. This will not affect our discussion
here.}
Furthermore, when two sets of
D-particles are separated  in a direction parallel
to the D8-brane,  cluster decomposition implies  that $m$ is
independent of the number of D-particles, $n$. We will identify $m$ with
 the  mass parameter of  the type-IIA  supergravity \cite{romans}  in the
region to  the right of the D8-brane \cite{joe,bgpt}.

Let us  now focus  on a single D-particle  in the Born-Oppenheimer
approximation where its  position  is  slowly varying.   The
canonical anti-commutation relation for  the `fast mode',
\begin{equation}
 \{\chi, \chi^\dagger\} = 1 \ \ ,
\end{equation}
can be  realized in a two-state space ---  the `vacuum' defined by
$\chi \vert\; 0> = 0$,
and the `one-string'   state $\chi^\dagger \vert\; 0 > = \vert\; 1>$.
The difference in energy,  $ E_1 -E_0  = {g} Y_9$,
 is positive when the
particle is on the right of the D8-brane and can be interpreted
as the energy of an (oriented)  string stretching
leftwards  from the particle  to  the
brane.  However, as the particle crosses to the other side of the
D8-brane   the roles of
the two states are  exchanged: $\vert\; 1>$ becomes  the new ground
state of the system, while
$\vert\; 0>$ should  be interpreted as
a state with an oriented 
string stretching leftwards  from the D8-brane to the particle.
Therefore,  a
fundamental string is created or destroyed when a D-particle is moved
adiabatically through a  D8-brane.    What
makes this
phenomenon inevitable   is the fact that
both states  cannot be  physical at the same time.
Gauss' constraint,
\begin{equation}\label{gausscon}
(\chi^\dagger\chi + m)\ \vert\; {\rm phys}> = 0 \ ,
\end{equation}
is satisfied by the state $\vert\; 0>$  if $m=0$, and by the  state
$\vert\; 1>$ if $m=-1$. In either case the physical state is unique
and cannot therefore change during the process.

In either case, furthermore, the
 physical D-particle feels no (velocity-independent) force as it
moves slowly through the D8-brane. This follows rather trivially from
the expression for the lagrangian, (\ref{lagr}). 
Because of (8,0) supersymmetry,
the position $Y^9$ only occurs  in the
combination $(A_0 + Y^9)$ so that 
Gauss' law  $\partial {\cal L}'/\partial A_0
=0$ implies that the force on the  D-particle also vanishes.
 Physically, we may
interpret this result as coming from the cancellation of  two competing
effects, both  linear in the
displacement  $Y^9$:  as the D-particle moves leftwards through the
D8-brane the mass of the type IIA supergravity jumps by one unit,
and there is a discontinuous change in the slope of the effective  
inverse string coupling constant.
The D-particle, whose effective mass is proportional
to the  inverse coupling, therefore  feels a
 force that apparently  jumps 
discontinuously.  However, this discontinuity in the force  is  balanced precisely by the
tension of the fundamental  string  which is created or destroyed 
in the process \cite{klebanov,bgl}.

For  $m\not= 0, -1$  neither  of the two states of $\chi$ is physical.
This  apparent inconsistency  is  an artifact of the quantum-mechanical
truncation of the system.
The full string theory contains
excited  $(08)$  strings,  which can carry any  (positive or
negative) integer
multiple of the elementary $U(1)$ charge, so that any integer value
of $m$ is allowed.  
Gauss' law fixes the {\it net}  number of
strings oriented {\it towards} the   D-particle to equal $m$ (or $m+1$) in
the region to the right (left) of the D8-brane. 
The non-chiral excited strings can  flip their orientation as the
D-particle traverses the D8-brane, while the chiral 
$\chi$-string must be either created or  destroyed.
The origin of the arbitrary parameter $m$ can be traced in
 string theory  to the presence of
extra D8-branes located at $Y^9\to \pm \infty$. Long 
 strings joining   these distant 
D8-branes to the particle can break in the proximity of another D8-brane,
as we will see in some explicit examples in the sequel.

  The general phenomenon of brane creation, which is an important
ingredient of the gauge-engineering constructions of field theory \cite{hw,egk}, 
was viewed  in \cite{bdg}
as the T-dual of the  anomaly inflow argument \cite{callan} applied
to the intersection domain of a pair of branes \cite{ghm}.
A  similar reasoning  clarifies   the other `mysterious'
rule of gauge engineering, the empirical 
  `s-rule' of  Hanany and Witten \cite{hw}.
The  rule states that  a configuration with  more
 than one D3-brane joining a NS-fivebrane   and a D5-brane, and extending
in all space-time dimensions (an s-configuration), 
cannot have a quantum state of unbroken supersymmetry. 
This rule resolves an apparent  contradiction of string theory
 with the field-theoretic
fact that ($2+1)$-dimensional N=8 U(k)  gauge theory  with $k>1$ 
has no supersymmetric vacuum when one turns on a Fayet-Iliopoulos coupling. 
We will now  relate this s-rule to the Pauli-exclusion principle
for  the fermionic $\chi$-strings
of the previous section. A very different  argument for this rule
has been presented in ref. \cite{vo}.

The chain  of dualities transforming the system of a  D8-brane and a D-particle
  to the Hanany-Witten configuration \cite{klebanov,bgl}
 starts with three T-dualities
that map the D8-brane to a D5-brane and the D-particle to a D3-brane.
A further  S-duality maps the D5-brane to a NS five-brane, while leaving
the D3-brane invariant. Finally two extra T-dualities map this to
a  configuration
 containing a  NS five-brane   and a  D5-brane. The fundamental
(08) strings transform under this sequence of dualities to
D3-branes suspended between the two five-branes. The fact that only
a single (08) string can be in its supersymmetric fermionic ground
state translates  directly to the s-rule of Hanany and Witten.
Notice that when   the   transverse direction is 
compact  the rule does not forbid a BPS state with multiple  suspended branes of different winding numbers. 
\footnote{In this case the starting  D8-brane configuration  is
   inconsistent without the addition of orientifolds but  other
dual configurations, such as  two orthogonal D4-branes, for  which flux
can escape to infinity are  consistent.} 
Furthermore, there  is no  restriction on the number of (zero winding)  suspended branes
 if all but one of them are in excited non-BPS states.  Such 
configurations
break   supersymmetry but may nevertheless be stable since the Pauli principle
does not allow the excited branes to decay into their ground state.
At face value this accounts  for the absence of a supersymmetric field-theoretic vacuum  state 
in the $U(k)$ theories of   \cite{hw,egk,egkrs} when $k>1$, although 
 the decompactification limit of the T-dualized dimensions
 may be very subtle. 

\vskip 0.3cm

\section{D-particles in type I' string theory}

As shown in \cite{polwitt} the  orientifold projection on
nine-dimensional type
IIA superstring  theory that leads to the type I' theory introduces two
orientifold 8-planes.    In addition, the weakly-coupled type I'
vacuum state
contains sixteen parallel D8-branes together with their sixteen mirror
images in the
orientifold planes.    We will here generalize the earlier
discussion to take
into account the effect of this background on D-particle dynamics.

 In order to justify the restriction  to the quantum mechanical  
system we will  consider low-energy
processes in which the effects of excited and/or 
winding strings can be neglected. This can be achieved at weak
type I' coupling by  taking one orientifold plane together with   half the
 D8-branes to infinity in the ninth direction, 
while  leaving the first orientifold  with the
 remaining D8-branes within a sufficiently small (sub-stringy)
neighborhood of the origin.   However, as is usual in BPS situations,
  some of our
formulae will   continue to hold much beyond the range of parameters
in which they would have normally  been expected to be valid.
This will be confirmed explicitly in the following section,  where
we will compare our results to the  heterotic/type I
duality predictions.

\subsection{Effects of  an orientifold plane}

One  effect of  an  orientifold  is to project
onto   the symmetric parts of the supercoordinates $Y_j$
and $S_a$,
and the antisymmetric parts of $Y^9$, $S_{\dot a}$ and  $A_0$. To avoid
confusion we will denote  these new matrix-coordinates  by
\begin{equation}
X_j^{rs} =  Y_j^{\{rs\}} \ , \qquad
\Theta_a^{rs} =  S_a^{\{rs\}}
\end{equation}
and
\begin{equation}
\Phi^{rs} = i (Y^9)^{[rs]} \ , \qquad
V_0^{rs} = i A_0^{[rs]} \ , \qquad
\lambda_{\dot a}^{rs} =  i S_{\dot a}^{[rs]} \   ,
\end{equation}
where $\{\;\}$ and $[\;]$ denote symmetrization or
antisymmetrization.   The
world-line gauge symmetry is thus truncated from $U(n)$ to
$SO(n)$. The antisymmetric-matrix   coordinate $\Phi$ describes
 the displacements of {\it mirror pairs} of
D-particles in the direction transverse to the orientifold plane.
Finally the chiral fermions  $\chi_{I}^r$,  describing
 strings that stretch between  the D-particles  and  the  D8-branes,
 transform  in the  real
$(n,2N)$ representation of $SO(n)\times SO(2N)$, where $2N$ is the
number of
D8-branes (which will later be taken to be sixteen).

The full string calculation of the potential felt by   D-particles
in this background now includes non-orientable as well as orientable
world-sheets.   This means that, in addition to the $(88)$ and  
$(08)$ cylinder
diagrams that entered the IIA theory,    M\"obius strips must
also be included.  These describe the effect of the mirror images of the
D-particles and  implement the (anti)symmetrization described above.  As
before, the excited string states are non-BPS degrees of freedom
and decouple from the low-momentum processes that concern us here.   A
potentially important effect arises from  the fact that strings can  
wind around
the compact ninth dimension giving rise to a rich spectrum
 of BPS states.   Even though these winding states decouple in the limit
 of large separation of the orientifold planes and will be largely
 irrelevant in the following,  
it proves
very instructive to include them in the evaluation of the complete string
diagrams.  The effect of the winding numbers on the individual  
cylinder and
M\"obius strip diagrams is to introduce a {\it quadratic} dependence  
on the
positions of the D-particles, $Y^9$ (or $\Phi$) into
the potential.  Such a dependence is inconsistent with $(8,0)$  
supersymmetry
and, not surprisingly, cancels when all diagrams are added  
together.   This
cancellation provides a powerful  check on  the consistency of the  
calculation.

The   quantum mechanical lagrangian for this system  can be expressed as
\begin{equation}
\eqalign{
 {\cal L} &=  \ {1\over 2}  \;
 {\rm tr}\; \Bigl(   DX_j  DX_j + {g^2\over 2} [X_i,X_j]^2 -
D\Phi D\Phi
- {g^2}  [\Phi, X_i]^2
 \Bigr)
  + {i \over 2}\
 {\rm tr}\; \Bigl(
 \Theta_{ a}  D \Theta_{ a}  - g \Theta_{ a} [\Phi, \Theta_{ a}]
\cr &
-  \lambda_{\dot a} D \lambda_{\dot a}
 -  g \lambda_{\dot a} [\Phi,\lambda_{\dot a}]
+2g  X_i \gamma_{a\dot a}^i \{ \Theta_a ,
 \lambda_{\dot a}\} \Bigr)
  +\; {i}\; \Bigl(  \chi_I^{\ T}  D \chi_I
+  g\; \chi_I^{\ T}  \Phi \chi_I  + g m^{IJ}\; \chi_I^{\ T}  \chi_J
\Bigr)\ . \cr}
\label{fermlag}
\end{equation}
This  is the truncation of  the previous  lagrangian,
eqs. (\ref{lag}) and (\ref{lagr}), with two notable differences.
First  there is an extra term proportional to the
antisymmetric matrix  $m^{IJ}$, whose skew-symmetric
eigenvalues are  the
positions of the D8-branes, relative to the plane of the orientifold.
Secondly,  there is no Chern--Simons
term since the orthogonal gauge groups generally  have   no simple
$U(1)$ factors.
Such a term may seem possible in the special case $n=2$, but
it is actually forbidden by  the requirement of
cluster decomposition for  a system of
more than two D-particles.

 The hamiltonian,  Gauss constraint and supersymmetry algebra
can be derived similarly
with the result,
\begin{eqnarray}
 {\cal H} &=& \ {1\over 2}\;
{\rm tr}  \Bigl(  \Pi_j^2  -  \Pi_\phi^2
 + {g^2}  [\Phi, X_j]^2
- {g^2 \over 2} [X_i,X_j]^2 \Bigr)
+{i g\over 2} \;  {\rm tr} \;  \Bigl(
\lambda_{\dot a} [\Phi,\lambda_{\dot a}] +
 \Theta_{ a} [\Phi, \Theta_{ a}]\nonumber\\
 & & \qquad\qquad
-2 X_j \gamma_{a\dot a}^j \{ \Theta_a ,
 \lambda_{\dot a}\} \Bigr) -  {ig} \;
\Bigl(  \chi_I^{\ T} \Phi \chi_I
+\; m^{IJ}\;  \chi_I^{\ T}  \chi_J \Bigr) ,\label{hamdef}\\
{\cal G} &=&  [\Pi_j,  X_j]  -  [\Pi_\phi,  \Phi]
+i \Theta_a \Theta_{ a} -i  \lambda_{\dot a} \lambda_{\dot a}
+i \chi_I \chi_I^{\ T}\
\bigg\vert_{antisym}= 0  \label{gauss} \  ,
\end{eqnarray}
and
\begin{equation}\label{algebra}
\{ Q_{\dot a}, Q_{\dot b} \}\;  =\;  2 \delta_{{\dot a}{\dot b}}\;
\Bigl(  {\cal H}
 -  {g}\;  {\rm tr}( \Phi
 {\cal G}) +ig \; m^{IJ}\;  \chi_I^{\ T} \chi_J  \Bigr)\ ,
\label{suor}
\end{equation}
where the Gauss constraint is an antisymmetric-matrix equation.
The matrix momenta are  given by
 $\Pi_j = {\dot X}_j$, and $\Pi_\phi = {\dot\Phi}$.
For completeness we also give the canonical commutation relations,
suppressing the obvious $SO(8)$ indices,
\begin{eqnarray}
&&i [ \Pi^{rs}, X^{pq}]  = \{\Theta^{rs},\Theta^{pq} \}=
  {1\over 2}
 ( \delta^{rp} \delta^{sq} +  \delta^{rq} \delta^{sp} )\nonumber\\
&&i [ \Pi_\phi^{rs}, \Phi^{pq}]  = \{\lambda^{rs},\lambda^{pq} \}=
  {1\over 2}
 ( \delta^{rp} \delta^{sq} -   \delta^{rq} \delta^{sp}
)\label{comferms}\\
&&\{\chi^r_I, \chi^s_J \} =  {1\over2} \delta_{IJ}  \delta^{rs}  \ .
\nonumber
\end{eqnarray}

   An important point    is that
 $SO(n)$ gauge invariance,  together with the eight supersymmetries,
fix all   operator-ordering ambiguities of  this system.
Both the supercharges $Q_{\dot a}$ and the  Gauss-constraint   
operators, which
must satisfy the $SO(n)$ Lie algebra,  are
manifestly free
from  ordering ambiguities.  The supersymmetry algebra  (\ref{suor})
 then uniquely fixes  the order of operators
in the hamiltonian. The extra term proportional to
$m^{IJ}$ in this algebra is central  because
 the $\chi$-fermions commute with the
supercharges \cite{banksetal}. Note also that (8,0)
supersymmetry  again allows an
arbitrary  kinetic term in the $\phi$ direction. Under some mild
assumptions this will not
affect our discussion of  slowly-moving, semi-classical D-particles.

  The simplest case ($n=1$) corresponds to a single D-particle   
that is stuck
on the
orientifold
plane.   Since the particle is stuck its transverse
kinetic energy must vanish  and its
total mass  should be determined exactly by its potential energy.
In order to
diagonalize the  hamiltonian, we make a unitary change of basis
$\chi_I \equiv U_{IJ}\; \chi_J$, such that
\begin{equation}
(U m U^\dagger)_{IJ}\;  =\;  im_I \delta_{IJ}\;
= \;  {\rm diag}\; (im_1, ...i m_N, -im_1,... -im_N)  \ .
\end{equation}
We shall use the notation $I \equiv ({\cal I}, \overline{\cal I})$, where
${\cal I}=1,..,N$ labels eight independent
 D8-branes  and
$\overline{\cal I}$ labels
their mirror
reflections. This amounts to the  decomposition $2N= N\oplus
\overline N$ under
the $U(N)$
subgroup of $SO(2N)$, with the reality condition
 $ \chi_{\overline{\cal I}}= \chi_{\cal I}^\dagger $. The
hamiltonian, eq. (\ref{hamdef}),  of the
D-particle
takes the simple non-interacting form
\begin{equation}
{\cal H}^{(n=1)}\; = \; {1\over 2} \Pi_j^2 + gm_{\cal I}\;
[\chi^\dagger_{\cal
I},
\chi_{\cal I}]\  ,
\label{spec}
\end{equation}
 with implicit summation on  the index ${\cal I}$ and with
\begin{equation}\label{anticom}
\{\chi^\dagger_{\cal I},
\chi_{\cal J} \}=  {1\over 2} \delta_{{\cal I}{ \cal J}}.
\end{equation}
The total  mass of the D-particle is then given by
\begin{equation}\label{spinmass}
M = M_0 + gm_{\cal I} \; q_{\cal I},
\end{equation}
where $M_0$ is the `bare' rest mass of D-particles
 and $q_{\cal I} = (\pm{1\over 2}...
\pm{1\over 2})$ are the weights of the  $SO(2N)$ spinor,  which
realizes the
fermionic anitcommuation relations. A discrete $Z_2$ remnant of the
local symmetry, which changes the sign of the $\chi$'s,  forces this
spinor to have definite chirality.
If $2{\tilde N}$
D8-branes coincide with the orientifold, the
stuck D-particle  will describe a  degenerate spinor representation  of
$SO(2{\tilde N})$.
By realizing  the algebra of the $\Theta_a$
zero modes, the  particle also carries   a vector-supermultiplet
representation of the spatial  $SO(8)$.

The case $N=8$ is special because in this case the theory in the bulk
is massless type IIA supergravity with freely-propagating gravitons.
The bare mass $M_0$ of D-particles, which are Kaluza-Klein supergravitons,   
is therefore well defined far from the D8-branes.
Since the mass difference $M- M_0$
 in (\ref{spinmass}), which  depends linearly   on the
positions of the D8-branes,   can be negative,  
there exist 
sub-threshold bound states.    Consider  a specific example that
will be of
relevance in the next section in which seven  D8-branes and
their mirrors
coincide with an  orientifold plane  ($m_{{\cal I}} =0)$ with a  
single mirror
pair
located at $\pm
m_1$.   The space-time  gauge symmetry associated with this
configuration
is $SO(14) \times U(1)$.    From (\ref{spinmass}) it follows that
the states
divide into two groups with positive and negative binding energies
and so they
have  masses
\begin{equation}
\label{twogroup}
M^\pm = M_0 \pm {1\over 2} g m_1.
\label{spinorsplit}
\end{equation}
There are therefore $2^6$ degenerate sub-threshold bound states in this
configuration which fill out  a $({\bf 64}, -{1\over 2})$  chiral spinor
representation of $SO(14) \times U(1)$.
Similarly, there is a degenerate state consisting of an anti  
D-particle stuck
on the orientifold plane that describes a $({\bf 64'},{1\over 2} )$  
spinor of the
opposite $SO(14)$ chirality. 
Two D-particles in such a state would not be able to leave the  
orientifold
plane  even if they happened  to collide.


\subsection{Bound states of a mirror pair}

In order to further elucidate the mechanism responsible for binding 
the D-particles to the orientifold, we
 turn now  to  the case $n=2$ which allows for motion of a mirror pair
transverse to the orientifold plane. We are interested in the 
 `Coulomb'  branch along which $\phi$ is non-zero. 
 We will work within a Born-Oppenheimer
approximation, and focus attention on the massive `fast'  
coordinates, which are
those charged under the $SO(2)\simeq$
$U(1)$ gauge symmetry on the world-line. 
 The charges of the various fields are as follows.
Antisymmetric matrices
are  neutral,   vectors   have  charge equal to one, while  symmetric
matrices contain  a neutral trace part  and a charge-two complex
component.   Accordingly,  the symmetric matrices may be replaced by the
charged creation and
annihilation operators,
\begin{eqnarray}\label{}
&&{1\over 2}( {X^{22}_j-X^{11}_j}) +
  i X^{12}_j ={1\over \sqrt{2}} (a_j + {b_j}^\dagger)\nonumber\\
&&{1\over 2} (\Pi^{22}_j- \Pi^{11}_j)
 - i \Pi^{12}_j = {i \over \sqrt{2}}  ({a_j}^\dagger -  b_j) \\
&&{1\over 2} ( \Theta^{22}_{ a} -\Theta^{11}_{ a}) +
  i \Theta^{12}_{ a} =   \theta_{ a}\nonumber  ,
\label{redefs}
\end{eqnarray}
and the (08) fermions may be complexified,
\begin{equation}
   \chi_I^1 + i \chi_I^2 \equiv \chi_I \ .
\end{equation}
The complex fermions, $\chi_{\cal I}$ and $\chi_{\overline{\cal  
I}}$, are now
independent and have the same  (negative)  U(1) charge.
The canonical anticommutation relations are
\begin{equation}
[a_i,{a_j}^\dagger]= [b_i,{b_j}^\dagger]= \delta_{ij}\ \ , \ \ \
\{
{ \theta_{ a}}, \theta_{ b}^\dagger \}=\delta_{ a b} \ \  ,
\ \  \{\chi_I, \chi_J^\dagger\} = \delta_{IJ}\ .
\end{equation}

 The string interpretation of these operators is exhibited  in figure 1.
The operators $a_j^\dagger$ and $\theta_a^\dagger$ create
bosonic and fermionic
strings joining the D-particle to its mirror image, and  oriented
{\it towards} the D-particles.
 The operator $b_j^\dagger$ creates a  similar bosonic string, but
oriented {\it away} from the D-particle and its image.  Finally,  
the operator
 $\chi_I^\dagger$ creates a  fermionic
string  stretching {\it away} from  the D-particle
  towards  the Ith D8-brane  if the latter lies between the
particle  and its mirror image.  Otherwise,  $\chi_I^\dagger$  
annihilates a
string stretching  from  the D8-brane {\it towards} the D-particle.
The orientation of these strings reflects their chirality in the  
dual type I
description, and  fixes  the sign of their U(1)  charge.
Gauss'  constraint, eq. (\ref{gauss}),  reads
\begin{equation}
 {\cal G}_{21}\;   =\;      a_j^\dagger a_j   +
    \theta_a^\dagger \theta_a -   b_j^\dagger b_j
   - {1\over 2} \chi_I^\dagger  \chi_I  - (4-{N\over 2}) \;
 =\;  0\ .
\label{GG}
\end{equation}
The U(1) charges of the various strings,
+1, +1,   $-$1  and
$-$1/2, are read off in  the order they appear. A mnemonic for Gauss' law
is that it fixes the net number  of arrows pointing  {\it towards}
the D-particle {\it or}  its mirror image.  The mirror images of  
$\chi$-strings
do not contribute  to  this counting,  nor to the total energy, and  
are thus
drawn with
faint lines  in the figure.

\vskip 0.4cm

\ifig\fone{ The four types of stretched strings along  the  Coulomb  
branch in
the n=2 case, as  described in the text. Figures (a) and (b) depict  
the strings
stretching between the D-particle and its mirror, while (c) those between
the particle and a  D8-brane. The broken line is the orientifold.
 The mirror image  of the string  in (c) does  not
contribute to the charge and energy, and is   drawn as a faint line.
For a given point in the moduli space  Gauss'
law fixes the net number of arrows pointing towards  the particle or  its
mirror image.
The  energy of these  configurations
  is proportional to the total (non-faint) string length.}
{\epsfbox{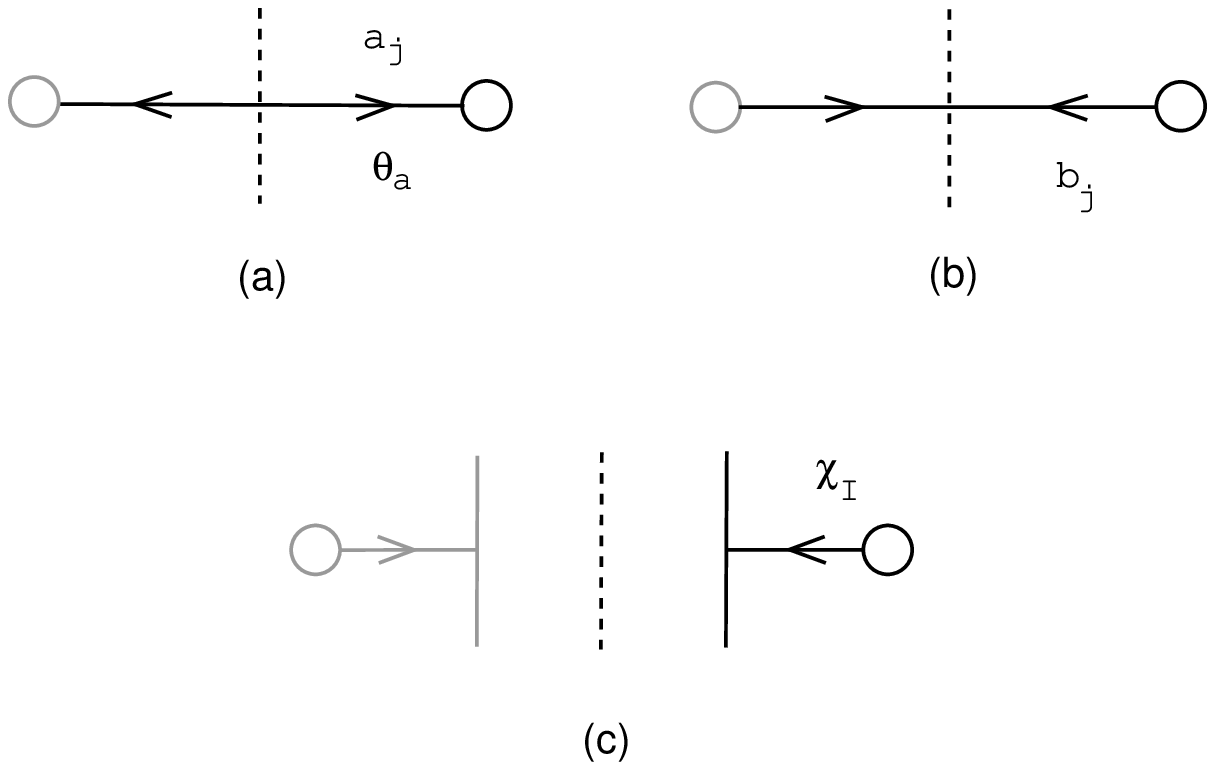}}

\vskip 0.4cm

We are now ready to discuss the dynamics of the mirror pair of
D-particles.
Consider, without loss of generality, the point $\Phi_{21} = \phi  
>0$ in moduli
space. Neglecting interaction terms, and performing a
$\phi$-dependent rescaling of the   bosonic matrix coordinates, allows us
to express the hamiltonian of `fast' modes as
\begin{equation}
{\cal H}_{\rm fast}^{n=2} \simeq
2g \phi\;
 (a_j^\dagger a_j + \theta_a^\dagger\theta_a+ b_j^\dagger b_j)
+ (\phi -  m_I) \chi_I^\dagger  \chi_I +  g\phi\; (8-N)\ .
 \label{ham}
\end{equation}
It is important that the order of operators, and the related
subtraction term,  is   fixed unambiguously
in the expressions for ${\cal G}$ and ${\cal H}$.
In the type IIA theory  we had the freedom to
choose  the  value of the mass parameter  $m$
in the far right asymptotic region,
by placing  a number of   D8-branes
at  infinity. Since the mass
jumped  precisely by one unit at each D8-brane, its value everywhere
else was fixed. In the present situation there is no ambiguity
whatsoever, because the value of the mass is fixed uniquely
 in the region
between the orientifold and the closest (mirror pair of) D8-branes.
This is the physical interpretation  of the fact that
$SO(n)$ quantum mechanics does not allow for  a Chern-Simons term.

 In   the region  to
the right  of all D8-branes the lowest eigenstate of ${\cal  
H}_{fast}$
is the naive  `vacuum'
\begin{equation}
( a_j\; , b_j\; , \theta_a\; ,\ {\rm  and }  \   \chi_I)\; \vert 0>  
= 0\ .
\end{equation}
It satisfies Gauss' law if and  only if there are precisely
$N=8$  D8-branes (and their mirrors),
 in which case  ${\cal H}_{fast} \vert 0> = 0$.
This is consistent  with the well-known fact \cite{polwitt}    that
the dilaton tadpole  cancels between an orientifold and eight
D8-branes, so that
 there is no dilaton gradient in the asymptotic region.
  Therefore, a distant
D-particle has  `no strings  attached' to it  and feels no static
force.
If all D8-branes coincide with  the orientifold,
 any bound states of the  mirror pair of particles
must be   threshold bound states.
Heterotic/type-I'
duality  predicts the
existence of such  bound states  \cite{KS,Lowe}, but this  fact  is
 hard to establish independently.

     The problem becomes simpler if  at least one   pair of D8-branes
moves  away from  the orientifold, in which case our  discussion
of stuck isolated D-particles suggest the existence of sub-threshold
bound states.  As a particular explicit example
 consider again
the  configuration  in which all  the D8-branes coincide with the
orientifold plane apart from one pair ($m_1 >  0$,  $m_{2...8}\simeq 0$).
This breaks the $SO(16)$ gauge group associated with  $(88)$ open  
strings  to
$SO(14)\times U(1)$.   It is  energetically-favourable for a (pair of)
D-particles
at position $\phi> m_1$  to be in the state
$\vert 0>$  with `no strings attached' (figure 2a).
As  it   crosses  the D8-brane  a string  is     anomalously created
( figure 2b).  However, in contrast to the earlier discussion
 of an  isolated D8-brane,
the quantum mechanics   now  has a richer spectrum of allowed states.
These include the bosonic and fermionic states,
\begin{equation}
\vert j\;I \rangle  =\;  a^\dagger_j\;
 \chi^\dagger_1 \chi^\dagger_I\; \vert 0\rangle \ , \ \ \ {\rm and}\ \ \
\vert a \;I \rangle  =
\; \theta^\dagger_a \chi^\dagger_1 \chi^\dagger_I\; \vert 0\rangle \ ,
\label{states}
\end{equation}
with $I=2,...8$ or $10,...16$.
These are obtained by trading the string attaching
the D-particle to the right-most D8-brane
for two strings ---  one  that attaches  it to its mirror image,  and
the other  to one of the  D8-branes  on  its  left (figure 2c).
\footnote{Positive-chirality strings stretching away from the orientifold
transform under the (8,0) supersymmetries, so that
at  $\phi\not= 0$ these states are not supersymmetric. This is consistent
with the fact that there is a net force pushing the D-particles to the
orientifold plane.}
The net number of arrows pointing to the D-particle and to its image
 is indeed conserved,
in agreement with our mnemonic for   Gauss' law. Put differently,
these states can be created from the `vacuum' by acting with the
 $SO(n)$-invariant operators
\begin{equation}
  V_{1I}^j
\sim \; {\rm tr} ( X^j \chi_1  \chi_I^{\ T}) \ \ {\rm or } \ \
V_{1I}^a \sim\; {\rm tr}(  \theta^a  \chi_1\chi_I^{\ T}) \ ,
\label{vertex}
\end{equation}
where the $\chi$'s are here $SO(n)$ vectors. This
 guarantees their consistency with the Gauss constraint.
The  quantum numbers of the
 states (\ref{states})  are those of   a  ten-dimensional vector
supermultiplet, in the ({\bf 14},$-1$) representation of  the
$SO(14)\times U(1)$ target-space gauge group.

\ifig\ftwo{A  D-particle pair in the background
of  an orientifold plane (broken line) and eight  D8-brane pairs, as
described in the text. Seven D8-brane pairs and their mirror pairs  (bold and faint thick lines) sit
at the orientifold, while one pair is moved out to position $\pm m_1$.
 In (a) the particle lies to the right of all D8-branes
and has no strings attached. Crossing the outermost
  D8-brane as in (b) leads to the
creation of a string, in accordance with Gauss' law. In (c) it has become
energetically favourable to trade this string for two others, one
stretching
to a D8-brane on the left, and one stretching to the particle's image.
The latter carries two units of charge, so the total charge of the state
is conserved.  }
{\epsfbox{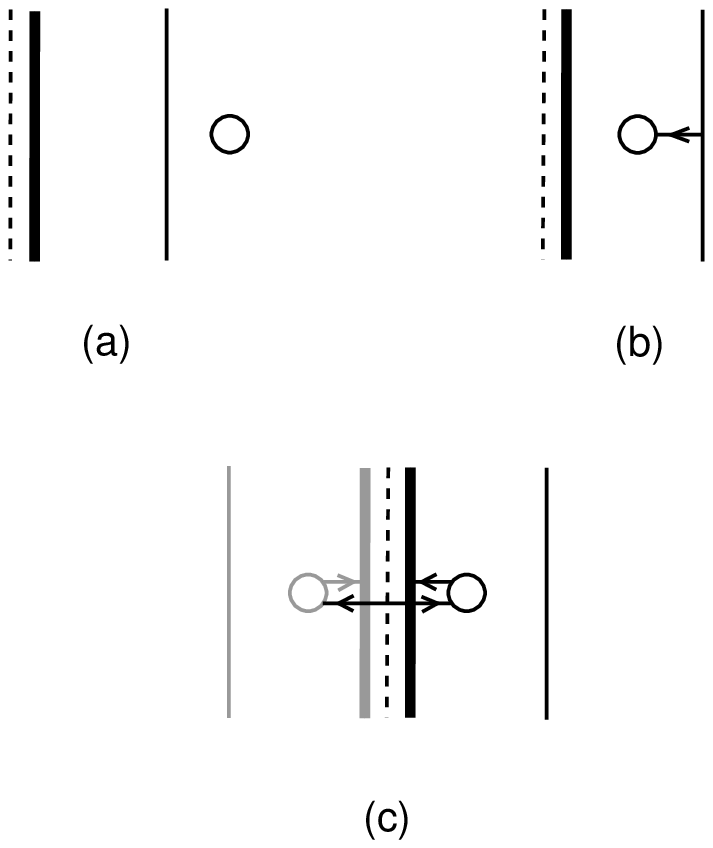}}

 The energy of the fast modes in any of these
states,    $E_{fast}  = g (4\phi - m_1)$,   is proportional to the
length of stretched strings in figure 2c minus the length of the
string in figure 2b, which was annihilated by the action of the above
operators on the naive vacuum. A (pair of) D-particles in one of these
states will therefore feel a constant force attracting it to
the orientifold plane.
 To leading order at  weak
string coupling, the mass of the resulting bound state is
 the total rest mass of the
D-particles  minus the energy at the bottom of the potential well,
\begin{equation}\label{boundmass}
M \simeq 2M_0 - gm_1\ .
\label{vectorsplit}
\end{equation}
Corrections to this formula may come from the neglected interaction  
terms,
the kinetic energy of the D-particles, and other higher-dimensional  
operators.
However,   the fact that the mass in (\ref{boundmass}) is precisely  
twice the
mass of the spinor states in   (\ref{twogroup}) suggests
 that these subleading effects miraculously cancel.  This  
observation is based
on  comparison with the heterotic string where it is easy to see  
(as we will in
the next section)  that the $SO(14) \times U(1)$  vector states are  
precisely
twice as massive as the spinor states.

The transition from the naive vacuum to one of the    states  
(\ref{states}) involves the emission of a massive $SO(16)$  gauge boson,
represented by an open string stretching
between the D8-brane on the right, and one of the fourteen  
branes
at the orientifold.   This  is a local interaction that occurs when the D-particle meets the orientifold plane.   It is associated with the operators  $V_{1I}^j$ and $ V_{1I}^a$, which are the vertex operators for the emission of such massive (super)gauge  
bosons in the
heterotic matrix model.
There is  however one further  subtlety since the  
quantum-mechanical model also
contains the fermionic zero-modes $\lambda_{\dot a}$
and ${\rm tr}\Theta_a$, whose Clifford algebra is realized by a
256-dimensional representation of the Lorentz group.
It would seem that this representation is carried by the `vacuum'  
$\vert 0>$,
 consistent
with the M-theory interpretation of D-particles as massive Kaluza-Klein
supergravitons in the bulk of the eleven-dimensional world  
\cite{horwitt}.
Does this mean  that the above bound states   transform in
a $16\times 256$-dimensional  representation of the Lorentz group?
The answer is no, because the D-particles interact with  massless
supergravitons, which are the massless closed strings in the bulk.
They  can carry
away  even-tensor  representations  of the transverse $SO(8)$
 without changing the semi-classical energy  of the system.
The same is true for the
massless $SO(14)$  gauge bosons,  which can be emitted or absorbed
 freely when the D-particles coincide with the orientifold plane.
Since these  effects cannot be accounted for
within the  Born-Oppenheimer treatment  of the system the quantum
numbers of the bound states can be only fixed modulo emission of
massless supergravitons  and massless super-gauge bosons.

One important
fact  should be stressed.  What we have established
is the existence of states binding the mirror pair of
D-particles to the orientifold plane.  However,  the two   particles  
could still escape
to infinity along the orientifold directions $X_j\sim x_j \sigma_3$,
i.e. along  the Higgs branch of the corresponding moduli space.
Comparing the mass formulae
(\ref{spinorsplit}) and (\ref{vectorsplit}) shows that  the    
vector state is
stable against this   decay  only  if  there is a threshold  bound  
state of
two D-particles stuck on the orientifold.  The existence of this  
threshold
bound state  will follow from target-space gauge invariance as
will be explained in the following section.

\subsection{Moving more D8-branes}

 First however we will  consider  the more
general situation, in which  more than one pair of D8-branes
 is moved off the orientifold plane.
To start with let us move two (pairs of)  D8-branes to  positions
$m_1$ and $m_2$, leaving all others  at the orientifold.
The open-string gauge symmetry is broken  down to
$SO(12)\times U(1)\times U(1)$. A single D-particle stuck at the
orientifold plane will carry a $SO(12)$ spinor
representation, and have a spectrum of masses
\begin{equation}
M = M_0\pm{1\over 2}gm_1 \pm {1\over 2}gm_2 \ .
\end{equation}
The lightest stuck D-particles will therefore be in the representation
$({\bf 32}, {\bf -{1/ 2}},{\bf -{1/ 2}})$.

  The binding mechanism of a  mirror pair  in this background is
illustrated in figure 3. When the particle lies to the right of the
D8-branes it has no strings attached to it as usual (figure 3a).
As the D-particle moves to the left,  crossing the two D8-branes,  
two strings
oriented towards the D-particle are created (figure 3b).
The total energy of the fast modes in this configuration is still zero.
The physical states of lowest energy, for sufficiently small $\phi$, are
now
\begin{equation}
\vert j \rangle  =\;  a^\dagger_j\;
 \chi^\dagger_1 \chi^\dagger_2\; \vert 0\rangle \ , \ \ \ {\rm and}\ \ \
\vert a  \rangle  =
\; \theta^\dagger_a \chi^\dagger_1 \chi^\dagger_2\; \vert 0\rangle \ .
\label{statestwo}
\end{equation}
They are obtained by trading the two strings attaching the D-particle
to the D8-branes on its right for  a single string attaching it to
the particle's
  image in the mirror (figure 3c). The energy of the fast modes in this
configuration is $E_{fast} = g( 4\phi-m_1-m_2)$, so the rest mass of a
bound state to (sub)leading order  at weak coupling  is
\begin{equation}
M \simeq  2M_0 - gm_1 - gm_2\ .
\end{equation}
This is again twice the   mass of the lowest-mass state of   a single
D-particle  that coincides with the orientifold  plane  
(\ref{spinorsplit}).
Comparison with the heterotic string in the next section will again  
indicate
that the approximate expression for the mass of the two D-particle  
state is  exact.

 States created by $V_{1I}^j$, $V_{2I}^j$ ($I\not= 1,2$)
 and their space-time spinor partners
are also lower in energy
than the naive vacuum $\vert 0>$
near the orientifold plane. They are reached by trading only one of
the two strings that attach the particle to a D8-brane on its right.
However, such states have  higher
mass than the states  (\ref{statestwo}), and can indeed decay
into them by emitting one  massive and one massless open-string
gauge boson at threshold. All states with $n>2$ have also similar
potential instabilities at threshold. Strictly-speaking,
the only conclusions that follow rigorously from our discussion in this
section are  (i) that the
lowest-mass n=1  states  are stable, and 
(ii) that some of the  n=2  states cannot decay by emission of a Kaluza-Klein
supergraviton. 

\vskip 0.6cm

\ifig\fthree{A mirror  pair of D-particles in the background with
two displaced D8-branes. In (a) the particle is to the right of the
D8-branes and has no strings attached. In (b) it has crossed to the
left, and two strings attaching it to the D8-branes have been
created. In (c) these have been traded for a single string attaching
the D-particle to its mirror image.
 }
{\epsfbox{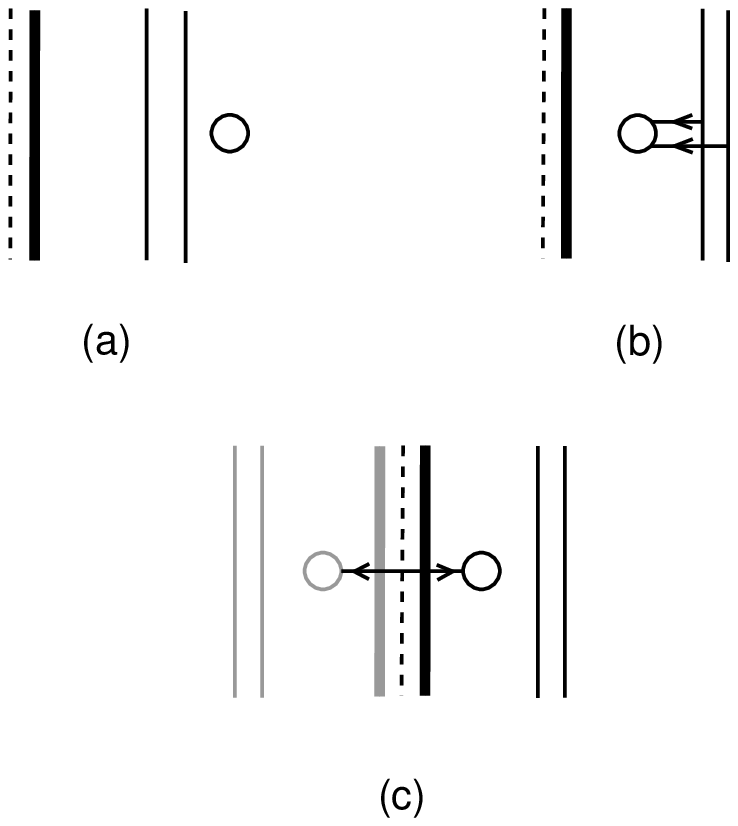}}

\vskip 0.3cm

 Situations where   more than two D8-branes are displaced from the orientifold
plane can be analyzed similarly. One novel subtlety,  that will play a role
later on,  appears when   all eight D8-branes are  displaced. 
The n=1  mass spectrum  (\ref{spec})  depends in this case
on whether the number
of positive $m_{\cal I}$'s  is even or odd.
Put differently, because the D-particle carries a chiral spinor 
representation of $SO(16)$, only an even number of D8-brane  reflections
in the mirror   leave its   spectrum invariant.
In particular, there are two inequivalent configurations with eight
D8-branes at position $m$, and their eight mirror images at $-m$. In
one configuration there is a unique
 lowest-lying D-particle state with mass  $M = M_0 -4gm$,
 in the $({\bf 1}, -4)$ representation of the $SU(8)\times U(1)$
open-string gauge group.\footnote{Our  normalization is such that a string
attached to one of the eight D8-branes has U(1) charge equal to $\pm 1$.}
 In the other  configuration, the 
lowest-lying state of the D-particle
is in a degenerate $({\bf 8},-3)$ representation, and
has   mass $M= M_0 - 3gm$.  This degeneracy 
can be pictured  as  the different ways in which  a string  can join any of the eight displaced  D8-branes to the D-particle (or anti D-particle). We will use this second
configuration in the discussion
of the special locus of  moduli space with $SU(18)$ gauge symmetry.

\vskip 0.3cm

\section{Global aspects of  moduli space}

A significant test of our understanding of  issues concerning
bound states in
the type I' theory  is whether we can reproduce the spectrum of BPS
states in
regions of  moduli space that are outside of the weakly coupled
domain. For example, using the duality relations between type I'
string theory and the heterotic string it is easy to see that there  
must be points in moduli
space (at strong type  I' coupling) at which the symmetry is enhanced
\cite{polwitt}.  These correspond to the compactifications of the
heterotic
string on a circle where  the symmetry is enhanced at specific values of
the radii and of the background  Wilson lines.  It was shown in
\cite{polwitt}
that these enhanced symmetry points occur in the type I' language
precisely when the effective couplng constant diverges at one of the
orientifold planes.  However, this gave
no hint of the mechanism that generates the extra massless states
needed to enlarge the symmetry.
In this section we will see   how such  enhanced symmetry points can be exactly
determined by the D-particle $(8,0)$ quantum mechanical hamiltonian.

\subsection{Heterotic string and M-theory on the cylinder}

Before describing the type I' dynamics we shall review the
expectations  based
on duality with the heterotic string or $M$-theory compactified on
a cylinder
\cite{horwitt}  of
circumference  $2\pi r_{11} $  and length $\pi r_9$. This
compactification has two dual interpretations
(i) as weakly-coupled  type I' theory in the limit $r_{11}\to 0$, and 
(ii) as the weakly-coupled
heterotic theory in the limit $r_9 \to 0$. 
More generally the  compactification has
a moduli space of dimension eighteen. On the heterotic side 
the moduli are the coupling constant $\lambda_h$,
 the radius $r_{11}$ of the cylinder, 
and sixteen Wilson lines in the Cartan subalgebra of the gauge group.
On the type I' side they are the coupling constant $\lambda_{I^\prime}$,
 the separation 
$\pi r_9$ of the  orientifolds, and the positions of the sixteen
independent D8-branes between the two  orientifold planes.

In order to
 work out the  precise mapping  between these two parametrizations of 
moduli space, we first consider  the special
locus where the gauge symmetry is
 $SO(16)\times SO(16)\times U(1)\times U(1)$. We have already seen that
on the type I' side this   is  a privileged configuration because it corresponds to
sixteen D8-branes
sitting  precisely at each orientifold.
Therefore, the (effective) string coupling is  constant over
the entire interval  and the two length scales of the cylinder
can be  varied independently without encountering any phase transitions. 
Since a mirror pair of D-particles is identified with a freely-propagating
Kaluza-Klein graviton in the bulk, we have
\begin{equation}
2M_0 = {1\over r_{11}}\ .
\end{equation}
The theory also contains BPS states that are membranes stretching 
a number $w$  of times between the two orientifolds. If $w$ is odd
the corresponding type I' string must start and end at two different
  D8-branes sitting
at opposite ends of the cylinder, and therefore carries charge under both
$SO(16)$ factors of the gauge group.  If 
 $w$ is even the string  may (or may not)  close
onto itself, and consequently may (but need not)  be neutral. 
 The minimum mass of a stretched membrane is $2\pi^2
r_9 r_{11}  T_2$, where $T_2$ is the membrane tension. The ratio
of the  winding and Kaluza-Klein masses is an  important dimensionless
parameter  
\begin{equation}
c =  2\pi^2 r_9 r_{11}^2  T_2 \ .
\end{equation}
The BPS spectrum of the theory does not depend on  the third scale,  
 which   can be chosen either
as the gravitational coupling,   or as the mass of (non-BPS)  open-string
excitations (${1\over \sqrt{\alpha_{I^\prime}}}
\sim \sqrt{T_2 r_{11}}$).

  The two extra  $U(1)$ factors of the gauge group  are associated with 
two gauge bosons, $G_{\mu\,  11}$  and $ C^{(3)}_{\mu 9\; 11}$, which 
 originate from Kaluza--Klein compactification of the eleven-dimensional
metric and  the three-form antisymmetric tensor potential.
The metric component
$G_{\mu 9}$ is projected out of the spectrum by the compactification,
and does not give rise to an  additional U(1).
On  the type I'  side  these  fields are the RR 
one-form coupling to
D-particle charge  and the Neveu-Schwarz antisymmetric tensor
coupling  to winding along  the ninth direction. In the heterotic
theory, they correspond to the off-diagonal components of the
metric and of the Neveu-Schwarz
antisymmetric tensor, which   couple to momentum and
winding along the eleventh direction, respectively. Under
T-duality momentum
gets exchanged with winding   and D0-charge with D1-charge.

  Let us move on now to the heterotic side, and 
 parametrize first the  moduli space in terms of
a Wilson line  coupling  to  $E_8\times E_8$ charge.
Such a Wilson line  is   a sum of two orthogonal
eight-vectors, ${\bf A}={\bf A}_{\bf 1}\oplus{\bf A}_{\bf 2}$, one for
each $E_8$ factor of the gauge group.  
The states of the theory are defined with reference
 to the  Lorentzian lattice, $\Gamma^{1,17}$,    given by
\begin{equation}
\label{hetlatt}
(\;p_L\; \vert\; p_R\; ) =  \left(\; {{\tilde m}\over r_{11}} - {w r_{11}
\over \alpha^\prime_h} \;
\Big\vert \;
{{\tilde m}\over r_{11}} +   {wr_{11}\over \alpha^\prime_h}
\; ;\; \sqrt{2\over\alpha^\prime_h}( {\bf Q}+ w{\bf A}) \; \right)
\end{equation}
where
\begin{equation}
{\tilde m} =  m  - {\bf Q}\cdot {\bf A} - {w \over 2} {\bf A}\cdot  
{\bf A}\ ,
\end{equation}
with $m$ and $w$ the integer  momentum and winding numbers,
${\bf Q}={\bf Q}_{\bf 1}\oplus {\bf Q}_{\bf 2}$
 a sixteen-component vector in the $E_8\times E_8$ root lattice
$ \Gamma^8\oplus\Gamma^8$, 
 and $\alpha^\prime_h$ the heterotic Regge slope.
The $E_8$ lattice $\Gamma^8$ is generated in our conventions by
the vectors $\pm e_i \pm e_j$ and
$\sum_i \pm {1\over 2}e_i$ where $e_i$ form an orthonormal set.
For a BPS state there are no oscillator excitations in the left-moving
sector, so the mass of the state is
 \begin{equation}
\label{masshet}
{ M^2 =  p_L^2 }.
\end{equation}

 The Wilson line ${\bf A}$ takes values in a fundamental  
cell of the self-dual  $E_8\times E_8$ lattice. In order to make
contact with the privileged type I' background, we 
 will choose the center of the fundamental  cell to be  the point of
$SO(16)\times SO(16)\times U(1)^2$ symmetry, 
\begin{equation}
  {\bf A} = \left( 1\; (0)^7\; 1\; (0)^7\; \right)\ +{\bf a} .
\label{center}
\end{equation}
Recall that the root vectors corresponding to the ${\bf 120}$ of $SO(16)$
have integer  entries while those corresponding to the ${\bf 128} $
have half-integer entries.
As a result, at the special point ${\bf a}=0$, the momentum $m$ of
states in the $({\bf 120},{\bf 1})$ and  $({\bf 1},{\bf 120})$
 is shifted by an
integer,  while  the momentum of the $({\bf 128},{\bf 1})$ 
 and  $({\bf 1},{\bf 128})$
 is shifted by a  
half-integer. This breaks the 
 $E_8\times E_8\times U(1)^2$ symmetry  down
to $SO(16)\times SO(16)\times U(1)^2$.

More generally, substituting (\ref{center}) into (\ref{hetlatt}) leads
to the following alternative form for the Lorentzian lattice, 
\begin{equation}
(\; p_L\; \vert\;  p_R\; ) = 
 \left(\; 
 {n-2{\bf q}\cdot {\bf a} \over 2r_{11}}-w 
({r_{11}\over \alpha^\prime_h}  +{{\bf a}\cdot{\bf a}\over 2r_{11}})
\; \Bigg\vert \;
{n-2{\bf q}\cdot {\bf a} \over 2r_{11}}+ w 
({r_{11}\over \alpha^\prime_h}  - {{\bf a}\cdot{\bf a}\over 2r_{11}})
\; 
 ;\;  \sqrt{2\over\alpha^\prime_h}
 ({\bf  q} +w {\bf a})\; \right),
\label{newlatt}
\end{equation}
where  
\begin{equation}
{\bf q} \in  \cases{ \Gamma^8\oplus\Gamma^8 & if $w$ is even \cr
 & \cr
( 1 (0)^7\; 1 (0)^7)+\Gamma^8\oplus\Gamma^8 & if $w$ is odd , \cr}
\label{charges}
\end{equation}
and $n$ is even or odd  depending on whether
 we are considering a tensor or a spinor representation
of  the diagonal $SO(16)$. 
We can now match precisely heterotic and type I' BPS states for the
special  value  ${\bf a}=0$ by noting that
\begin{equation}
\alpha^\prime_h  = (2\pi^2  r_9 T_2)^{-1} \ .
\end{equation}
The heterotic momentum and winding numbers,  $n$ and $w$, must
be identified with the number of D-particles and the number of type I'
 strings stretching
between the two orientifold planes, in accordance with the fact that
they are  charges for  the M-theory fields
 $G_{\mu\,  11}$  and $ C^{(3)}_{\mu 9\; 11}$.
The vector ${\bf q}$ gives the  charges under  the open-string gauge group
 on the type I' side.
Note, for instance, that type I' 
 states with $n$ odd have at least one D-particle stuck at
the orientifold plane and carry a spinor representation of 
the diagonal $SO(16)$, in
agreement with the above heterotic spectrum.
Furthermore, type I' states with odd $w$ are  strings stretching between
two D8-branes at opposite ends of the cylinder  and  therefore carry 
one vector index of  each $SO(16)$, as well as 
 extra charges in the
 $E_8\times E_8$ lattice. This is again in agreement with the heterotic
spectrum (\ref{newlatt}). Note finally that 
 the BPS  spectrum of the heterotic string
does not depend on the string coupling $\lambda_h$, which sets the value of
 the Planck mass. Unlike the tension of   type I' strings, the
heterotic string tension cannot be varied, however,  independently, because
$r_{11}^2/\alpha_h^\prime = c $.

\subsection{Enhanced symmetry and phase transitions}


In order to make contact with the quantum mechanical system 
of  section 3, we would now like to go to the  limit in which the
orientifold planes are infinitely far apart. This is the limit in
which stretched membranes are much heavier than Kaluza-Klein excitations
and decouple, which is the case if
\begin{equation}
c \gg 1 \ , \ \ \ \ {\rm or \ \ equivalently}\ \ \ \ \ 
 r_{11} \gg \sqrt{\alpha^\prime_h}\ .
\label{lim}
\end{equation}
The states that survive in this limit on the heterotic side have
$w=0$ and ${\alpha_h^\prime}(p_R^2 - p_L^2) = 0$ or $4$.
These are precisely  the Kaluza-Klein excitations
of the supergravity and $E_8\times E_8$ super Yang-Mills theory,
compactified to nine dimensions on a  circle. All heterotic string
excitations decouple, and so do   all states charged
under both  $E_8$ factors of the gauge group. We may therefore
 restrict our discussion to one of these factors, which
translates in  type I'
language  to focusing attention on a single  orientifold plane.

  Strictly speaking the limit (\ref{lim}) does not by itself justify
the  quantum mechanical truncation of the theory. To stay at weak
type I' coupling, we should  approach this limit by taking $r_9\to\infty$
while keeping $r_{11}$ small, in units of the eleven-dimensional Planck
scale. Furthermore, we must  consider processes sufficiently close to
the orientifold plane, for  which open-string excitations may  be
neglected. As usual with BPS statements, these caveats will not be
necessary, and the  mass formulae of the previous section will extend
much beyond their naive  range of validity. The only important  limitation 
comes from  decoupling of the winding BPS states, which are described by
a two-dimensional field theory,  and this decoupling
is guaranteed by the condition (\ref{lim}).

 The moduli space of the heterotic theory in this limit can be
explored by turning on a Wilson line ${\bf a} =
(a_1 ... a_8)$, 
for  one of the $E_8$ factors of the gauge group. This Wilson line
can be confined to a fundamental cell of the $\Gamma^8$ lattice, since
two backgrounds  that differ by a lattice vector are gauge equivalent. 
A two-dimensional section of such a cell, centered around the point
of $SO(16)$ symmetry, and corresponding to a Wilson line of the form
$(a_1\; a_2\; (0)^6\;)$  
is shown in figure 4.
Using the heterotic  mass formula, $M^2 =  p_L^2$, with $w=0$ gives 
the following spectrum of BPS states,
\begin{equation}\label{masshet}
M = {1\over 2r_{11}}\;\vert n-q_{\cal I}a_{\cal I}\vert
\end{equation}
where  $n$ is even for the supergravitons and $SO(16)$-adjoint states,
and odd for the states in the spinor representation of $SO(16)$.
For special values of the Wilson lines, $a_{\cal I}$, new BPS massless states arise, signalling the standard symmetry enhancements of the heterotic string.

The heterotic mass formula (\ref{masshet}) agrees precisely   with the spectrum of states in the
 type I' picture   if
we identify the Wilson lines and D8-brane positions by the relation,
\begin{equation}\label{arel} 
gm_{\cal I}= a_{\cal I}/2r_{11}\ .
\end{equation}
The exactness of the heterotic expression  strongly suggests that
our quantum mechanical mass formula (\ref{spec})
is also exact. It can therefore
be continued to a region in which the binding energy of the D-particle
is of the same order as its bare rest mass, in which case the bound state can be massless. For such critical displacements,
the type I' gauge symmetry will be non-perturbatively enhanced. 
This is to be distinguished from the standard perturbative symmetry enhancement in type I' that arises when D8-branes coincide with each other or  with an orientifold.

\vskip 0.3cm

\ifig\ffour{The two-dimensional section of the fundamental cell
of moduli space  discussed in the text.
 It corresponds to heterotic Wilson lines of the form $(a_1\;a_2\; (0)^6)$,
 or, in type I' language, 
to the motion   of two D8-branes away from  the orientifold.
 The center of the cell is  
 the point of $SO(16)$ symmetry where the D8-branes sit at the orientifold.
The shaded region is a single cover of the moduli space, obtained by 
modding out with the Weyl symmetry, i.e. 
the  permutations of the D8-branes. The generic
$SO(12)\times U(1)^2$ symmetry in the interior of the cell
is enhanced at the fixed loci of the Weyl group, as well as
at the outer boundaries.
The former enhancement is produced in type I' language by
colliding D8-branes, while the latter by massless bound states
of D-particles with the orientifold. The type I' description cannot
be continued analytically beyond this outer boundary.
}
{\epsfbox{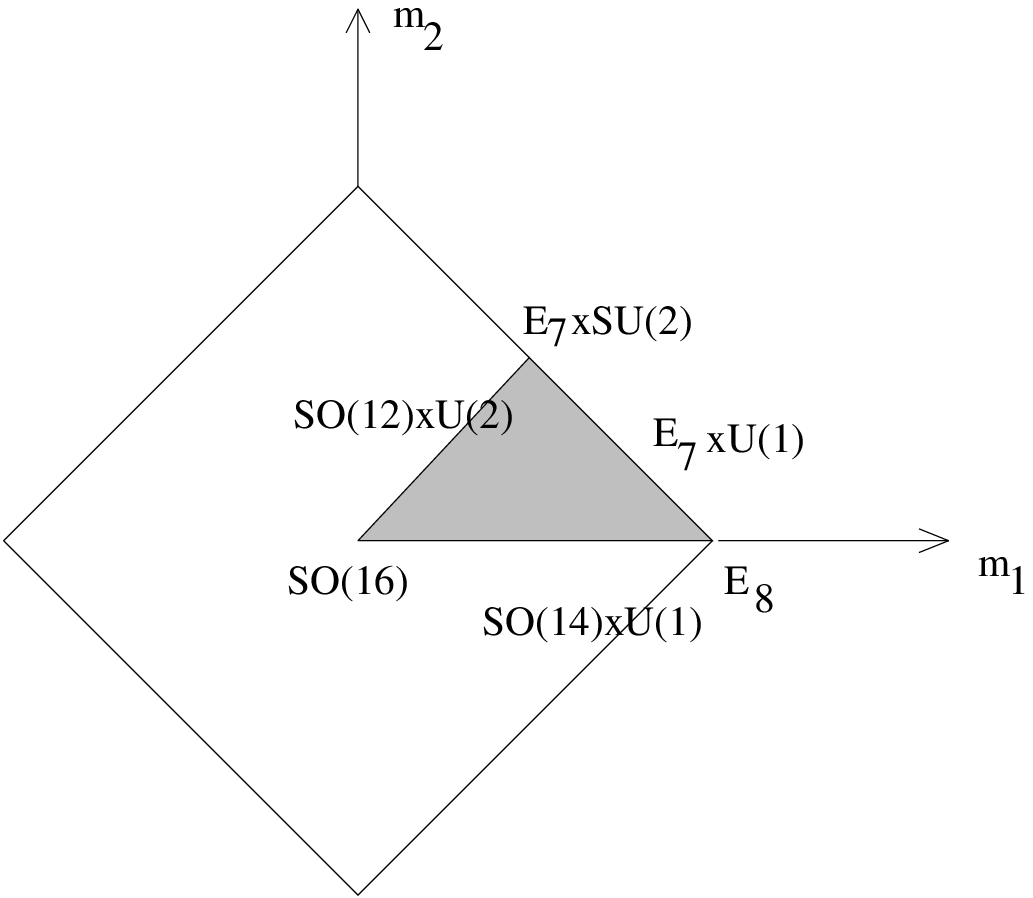}}

\vskip 0.3cm

It will be important in the following that all symmetry enhancements actually occur at the boundary of a single cover of moduli space. Such a cover,  which is represented by the shaded region in figure 4,   
 is obtained by modding out by the Weyl reflections of the fundamental cell.       These Weyl reflections 
 correspond to the permutation symmetries of the D8-branes. For instance
the reflection $m_1\to -m_1$ is simply the exchange of the first D8-brane
with its mirror image, together with an odd number of mirror exchanges of the D8-branes sitting at the orientifold.  The reflection $m_1\leftrightarrow m_2$
corresponds to interchanging the first two  mirror pairs of D8-branes. At a generic point in the interior of this region the symmetry is $SO(12) \times U(1)\times U(1)$.  
Along the fixed lines of the Weyl reflections --- the two interior boundaries of the shaded region   ---  the symmetry is enhanced in the standard perturbative  manner  while along the exterior boundary  it is   enhanced due to the appearance of  massless D-particle bound states.

We will now  describe  some of these points of enhanced
symmetry associated with a single orientifold in more detail. 
For example, we saw in section 3 that when a single mirror pair of D8-branes is displaced from an orientifold there are sub-threshold bound states of the symmetry $SO(14) \times U(1)$ associated with that orientifold.  These consisted of the 128 states in a non-chiral $SO(14)$ spinor and 28 states in a complex vector.   The masses of these states, given by  (\ref{twogroup}) and (\ref{boundmass}), vanish when the displacement of the D8-branes reaches the critical value $m_1 = 2M_0/g$.  This is a  point  of enhanced symmetry where the extra massless states combine with the 92 states of the adjoint of $SO(14) \times U(1)$ to form the ${\bf 248}$ of $E_8$.   
Moving two mirror pairs of D8-branes gives
an enhanced symmetry $E_7\times U(1)$ at a generic point
on the boundary of figure 4.
Generally, when $N$  D8-branes and their
mirrors are displaced, the unbroken
 $SO(16 -2N) \times U(1)^N$ gauge symmetry in the interior of the cell
is   enhanced to
$E_{9-N}\times U(1)^{N-1}$ at the boundary.
Here  the groups $E_1$,   $E_2$,
$E_3$, $E_4$, and
$E_5$ are conventionally defined as $SU(2)$, $SU(2) \times U(1)$,  
$SU(3)\times
SU(2)$,
$SU(5)$ and ${\rm Spin}\ 10$, respectively.  
One special point on the boundary of the fundamental cell corresponds to the heterotic Wilson line ${\bf a} = \left(\left({1\over 6}\right)^7\; - {1\over 6}\right)$ where the   symmetry is enhanced to $SU(9)$. In type I' language this arises when  all   D8-branes  are displaced to a critical distance from the orientifold in a such a way that the  $SU(8) \times U(1)$  perturbative symmetry is enhanced.   The extra massless states are the D-particle bound states in the $({\bf 8}, -3)$ representation of the perturbative group discussed at the end of section 3.

The semi-classical type I' description  cannot be
 analytically  continued beyond the boundary of the fundamental cell,
where  some D-particles have  binding energy that exceeds their  mass.
This is in contrast with the heterotic
theory (in the limit (\ref{lim})) which can be continued to  any point 
${\bf a}$.   However, since the Wilson lines
 have the periodicity of the $E_8$ root lattice, any such point
can be brought back to the fundamental cell provided one shifts
simultaneously the momentum  of the various states.   In the heterotic description the phase transitions arise when energy levels of fundamental quanta become negative, which is dealt with by a redefinition of the vacuum state.   In type I' language this would necessitate second-quantization of the D-particle bound states.   

The  picture of massless bound states  makes qualitative contact with the description of the type I' theory
based on  classical  supergravity  \cite{polwitt}.   
There the coupling  varies over space, 
and  diverges   at  one of the orientifold planes
when the symmetry on the heterotic side is
enhanced. 
 The  supergravity coupling is to be identified with 
the inverse of the {\it effective} mass of a D-particle,
 which varies as the particle moves through the potential.
  At an enhanced symmetry point the effective mass of a D-particle
 that is bound to an orientifold plane vanishes so that the effective
 coupling diverges  on that plane.  However, the description in terms
 of D-particle quantum  mechanics  captures the exact short-distance
 physics and allows precise calculations of the dynamics 
 at enhanced symmetry  points.

\subsection{Closing the orientifold gap}

 Our discussion in the previous subsection was confined to a region
of moduli space where the two orientifold planes were very far
apart. The physics of symmetry enhancement could thus be discussed
by focusing on one orientifold and ignoring the presence of the
other.  However, some of the 
interesting phenomena in heterotic string theory
  involve   winding heterotic $E_8\times E_8$
states, and necessitate the discussion of both orientifolds
 simultaneously.
This is the case,  for example, for  the special regions in moduli space where the
symmetry is  enhanced to $E_8\times E_8\times SU(2)$, $SU(18)$ or  $SO(34)$. These   only have  one free modulus,
 the heterotic coupling constant $\lambda_h$, and one may wonder
whether they can be attained in the type I' picture.
  Seiberg and Morrison
have in fact argued that  in order to describe some of these regions one may need
to introduce one or two extra D8-branes \cite{seibmor}. We will present some  persuasive arguments that  such regions can also be
described in terms of  massless D-particle bound states.   In order to give a complete discussion we would have to go beyond the quantum mechanical picture that applies when only one orientifold plane is present to a $(1+1)$-dimensional field theory description that takes into account the infinite tower of type I' open and closed strings  winding around the compact  ninth dimension.  
However, we will only present a qualitative  overview of the mechanism for symmetry enhancement.  
 
The  explicit examples of enhanced symmetries  to be considered below will make use  of  the following geometrical property of the  heterotic string that we believe to be true although we have not proved it.    Define 
 the moduli space of lorentzian $\Gamma^{1,17}$ lattices to be  ${\cal M}$  and  the subspace  in which  the gauge symmetry is larger  than
 $U(1)^{17}\times U(1)_g$ to be ${\cal M}_E$,   where the factor  $ U(1)_g$ denotes the abelian gravi-photon symmetry which can never be enhanced.     The property we shall use is that ${\cal M}- {\cal M}_E$ is {\it connected}.
 What makes  this  statement non-trivial  is the fact that some
regions  of enhanced symmetry  have codimension one, and
could conceivably separate moduli space into several disconnected components.
However, this is  not the case because such regions  are   boundaries, as illustrated in figure 4  for example,
 where all  symmetry enhancements occur at the
boundaries of the shaded region.\footnote{We thank K. Narain and  G.  Moore for  discussions
on this point.} 
Another simple example is the moduli space of $\Gamma^{2,2}$ lattices,
corresponding to compactifications of type II theory on a two-torus. If
$T$ and $U$ are the 
 complex structure and K\"ahler moduli of the torus, then all
 symmetry enhancements occur on the locus $T=U$. This is  again on
the boundary of moduli space because of the discrete symmetry
 $T\leftrightarrow U$.

  What this `theorem' implies in the cases we are interested in is that an
arbitrary  point in the  moduli space of interest can be reached from
 any other point continuously without ever encountering a 
phase transition. This guarantees that we may extend the  type I'
description to cover the entire moduli space of the theory, and reach, in particular,  the neighborhood of all the special points mentioned above,
without ever encountering negative-mass solitonic states.
  
For example, consider a type I' configuration with $E_8\times E_8\times U(1)\times U(1)_ g$ symmetry, obtained by 
displacing one (pair of) D8-branes to  a critical distance $m_{cr}
\sim (r_{11}^2 T_2)^{-1}$  from each
orientifold plane. There are two free parameters, the radii
$r_9$ and $r_{11}$, provided the latter is sufficiently small.
Decreasing the separation $\pi r_9$ of the
orientifolds, or equivalently the radius $r_{11}$ of the circle, 
brings the two separated  D8-branes closer. 
 At a critical value,
\begin{equation}
r_9 r_{11}^2 \sim T_2^{-1},
\end{equation}
 the two displaced  D8-branes  coincide (see figure 5). 
This   gives rise to an extra complex massless gauge boson, corresponding
to a stretched type I' string, and enhancing the $U(1)$ factor of the
gauge group to $SU(2)$. This  corresponds to  the straightforward compactification of the  heterotic $E_8\times E_8$ theory,
 with no Wilson lines, on a circle with self-dual
radius.

\vskip 0.3cm

\ifig\ffive{The type I' vacuum close to the point with $E_8\times E_8\times SU(2)\times
U(1)_g$ symmetry. It is obtained by first displacing the
two D8-branes to a critical distance from their respective orientifolds,
so as to enhance each $SO(14)\times U(1)$ to an $E_8$, then closing the
orientifold gap so as to make them collide.
}
{\epsfbox{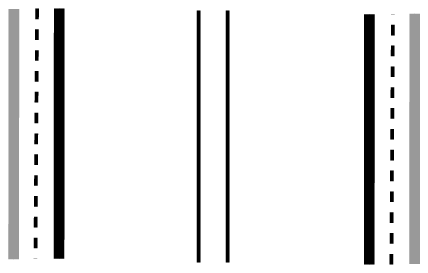}}

\vskip 0.3cm

A possible source of confusion is the identification of the $U(1)$ factors  that complete the three enhanced simple groups, two $E_8$'s and a $SU(2)$.  From the heterotic point of view, however,  their identification is clear.  At a generic point in moduli space the gauge field action has the form
\begin{equation}
\label{gaugefo}
{\cal L}_{gauge} = -{1\over 4} M^A_{\ C} \eta^{BC} F_A F_B,
\end{equation}
where $F_A$ ($A = 1, \dots, 18$) are the abelian  field strengths of the   $U(1)$ factors,  $\eta^{BC}$ is the $SO(1, 17)$-invariant metric and $M^A_{\ B}$ is a moduli-dependent Lorentz boost.  The   matrix $M $ describes the deformation of the Lorentzian lattice,  (\ref{newlatt}), as the Wilson lines
are turned on, 
\begin{equation}
\label{mdef}
\pmatrix{p_L \cr
           p_R\cr}_{\bf a} = M^{-1} \pmatrix{p_L \cr
           p_R\cr}_{{\bf a} = 0}.
\end{equation}
The perturbative open-string gauge group  
and the two extra bulk $U(1)$'s (the $C^{(3)}_{\mu911}$ and $G_{\mu \, 11}$ gauge potentials) that couple to Chan-Paton charges, D-particle number
and winding, respectively, do not mix at ${\bf a}=0$. However, as the
Wilson lines are turned on, or equivalently as the D8-branes are displaced, 
 the U(1)'s generally mix. 
 In particular, the charge vector of the $w=\pm 1$ string stretching
between the two displaced  D8-branes  is
 rotated in such a way that it becomes orthogonal to the charges of
the D-particle bound states at each orientifold plane, in accordance with
the fact that these states complete different simple factors of the enhanced
gauge group.

  In order to describe the  
 $SU(18)$ point  we start from  the region with  $SU(9) \times SU(9) \times U(1) \times U(1)_g$ symmetry  and with  the orientifolds  far apart.  As we saw earlier, this requires a configuration in which there is a   stack  of eight coincident D8-branes displaced a critical distance from each orientifold plane.  Reducing the separation of the orientifold planes as in the previous example again results in symmetry enhancement when the  two stacks coincide.  Among the massless states are the $U(16)$ open-string gauge bosons that arise in the standard manner.  Recall that each $SU(9)$ factor contains a massless D-particle bound state in a complex ${\bf 8}$ of  the perturbative  $U(8)$ subgroup which can  be pictured  as  the different ways in which  a string  can join  the D-particle (or anti D-particle) to any of the eight displaced  D8-branes.  Such a string is required to be present by the chirality argument given earlier.  When the two stacks of  D8-branes coincide the ${\bf 8}$ is augmented to a ${\bf 16}$ since the string can now terminate on any of the coincident branes.  Since there can be a D-particle on either orientifold plane there are two complex ${\bf 16}$'s giving a total of 64 extra massless states.  
  In addition, there is a complex $U(16)$ singlet  where one D-particle is stuck on each orientifold and they are joined by a $\chi$ string.   This gives a total of 66 states which completes the 258 states of the adjoint of $U(16) \times U(1) \times U(1)_g$ to  $SU(18) \times U(1)_g$.

Finally, we may consider the mechanism for the enhancement of the symmetry to $SO(34) \times U(1)$.  The point in moduli space where this occurs can   be reached  by starting from the  $SO(16) \times SO(16)$ configuration and    displacing   the stack of eight D8-branes (and their mirrors)  from the left orientifold plane.   When the coupling constant is sufficiently small  this  set of displaced D8-branes can be moved onto the right orientifold plane  without encountering a phase transition.     Now all 32 D8-branes coincide  and the theory is in a $SO(32)$  vacuum.   The degeneracy of the sub-threshold  ground states of a D-particle stuck on the left orientifold plane  is either 1 or 32 for the two inequivalent motions of the D8-branes. In one case the D-particle has no strings attached while in the other case (which differs by a single interchange of a D8-brane with its mirror) it is joined by a
single  $\chi$ string  to any of the 32 coincident D8-branes.\footnote{In heterotic language there are two inequivalent $SO(32)$ theories in nine dimensions that are distinguished only by a probe spinor.}   The coupling constant may now be tuned  so that in the second case   these 32 D-particle  states, together with the 32 states of an anti D-particle, give 64 new massless states.  These are precisely the 64 states that are needed to enhance the symmetry from $SO(32) \times U(1) \times U_g(1)$ to $SO(34) \times U_g(1)$.  The inequivalent $SO(32)$ theory with only two massless D-particle bound states is $SO(32) \times SU(2) \times U(1)_g$.
Other examples can be discussed similarly.
 
\vskip 0.3cm


\vskip 1.0cm

{\bf  Acknowledgements}

We thank G. Ferretti, M.  Gaberdiel,
I. Klebanov, K. Narain, G. Papadopoulos and  N. Seiberg  for
discussions.
We are also  grateful to  the organizers of the Newton Institute program on
{\it Non-perturbative Aspects of  Quantum Field Theory} for providing a
stimulating atmosphere during the early stages of this work.
This  work  was partially supported by  the EEC grants
CHRX-CT93-0340, TMR-ERBFMRXCT96-0090    and   by the 
Minerva Foundation,Germany.


\vspace{2cm}


\begin{thebibliography}{6666}

\bibitem{polwitt}  J.  Polchinski and E.  Witten, {\it Evidence for
Heterotic-Type I String Duality},  hep-th/9510169;   Nucl. Phys.
{\bf B460} (1996) 525.

\bibitem{horwitt}   P.  Horava and E.  Witten,
{\it  Heterotic and Type I String Dynamics from Eleven  Dimensions},
hep-th/9510209;  Nucl. Phys. {\bf B460}  (1996) 506.

\bibitem{DanFer} U.H. Danielsson and G. Ferretti, {\it The Heterotic
Life of the D-particle} hep-th/9610082; Int.J.Mod.Phys. {\bf A12} (1997) 4581.

\bibitem{banksetal} T.  Banks, N. Seiberg and E.  Silverstein, {\it
Zero and
One-dimensional Probes with $N=8$ Supersymmetry},  hep-th/9703052;
 Phys. Lett. {\bf B401} (1997) 30.


\bibitem{KS} S. Kachru and  E. Silverstein, {\it On Gauge Bosons in the
 Matrix Model Approach to M Theory}, hep-th/9612162;  Phys.Lett. {\bf B396} (1997) 70.

\bibitem{Lowe}  D.A. Lowe, {\it Bound States of Type I' D-particles and
 Enhanced Gauge Symmetry}, hep-th/9702006; Nucl. Phys. {\bf B501} (1997) 134.


\bibitem{Banmot} T. Banks and L. Motl, {\it Heterotic Strings from
Matrices}, hep-th/9703218.

\bibitem{rey} Soo-Jong Rey, {\it Heterotic M(atrix) Strings and
Their Interactions}, hep-th/9704158; Nucl. Phys. {\bf B502} (1997) 170.

\bibitem{hor} P. Horava, {\it Matrix Theory and Heterotic Strings
on Tori},
hep-th/9705055; Nucl. Phys. {\bf B505} (1997) 84.

\bibitem{KR} D. Kabat and S.-J. Rey, {\it Wilson Lines and T-Duality
in Heterotic Matrix Theory}, hep-th/9707099.

\bibitem{mall} D. Matalliotakis, H-P. Nilles and S. Theisen,
{\it Matching the BPS Spectra of Heterotic-Type I-Type I' Strings},
hep-th/9710247.


\bibitem{last} O. Bergman, M. R. Gaberdiel and  G. Lifschytz,
{\it  String Creation and Heterotic-Type I' Duality}, hep-th/9711098


\bibitem{bdg} C.P.  Bachas, M.R.  Douglas and M.B.  Green, {\it
Anomalous Creation of Branes},  hep-th/9705074;
JHEP07 (1997) 2.


\bibitem{klebanov}  U. Danielsson, G. Ferretti and  I.R. Klebanov, {\it
Creation of Fundamental Strings by Crossing D-branes},  hep-th/9705084;
Phys. Rev. Lett. {\bf 79} (1997) 1984.

\bibitem{bgl} O.  Bergman, M.  Gaberdiel and G.  Lifschytz, {\it Branes,
Orientifolds and the Creation of Elementary Strings}, hep-th/9705130.

\bibitem{alwis} S.P. de Alwis, {\it A Note on Brane Creation},
hep-th/9706142.

\bibitem{howu} P-M. Ho and  Y-S. Wu, {\it Brane Creation
 in M(atrix) Theory},  hep-th/9708137.

\bibitem{ohta} N. Ohta, T. Shimizu, J-G.  Zhou, {\it
Creation of Fundamental String in M(atrix) Theory},  hep-th/9710218.



\bibitem{hw}  A. Hanany and E. Witten,
 {\it Type-IIB Superstrings, BPS
Monopoles and Three-Dimensional Gauge Dynamics},  hep-th/9611230;
 Nucl.Phys. {\bf B492} (1997) 152.

\bibitem{egk} S. Elitzur, A. Giveon and  D. Kutasov, {\it
 Branes and N=1 Duality in String Theory}, hep-th/9702014;  Phys.Lett. {\bf B400}
 (1997) 269.

\bibitem{vo} H. Ooguri and  C. Vafa, {\it
 Geometry of N=1 Dualities in Four Dimensions}, hep-th/9702180;
 Nucl.Phys. {\bf B500} (1997) 62.

\bibitem{egkrs} S. Elitzur, A. Giveon, D. Kutasov, E. Rabinovici and
 A. Schwimmer, {\it Brane Dynamics and N=1 Supersymmetric Gauge Theory},
hep-th/9704104; Nucl. Phys. {\bf B505} (1997) 202.

 
\bibitem{Halp} M. Claudson and M. Halpern,  {\it Supersymmetric Ground State Wave Functions}, Nucl. Phys. {\bf B250}
(1985) 689.

\bibitem{deW} B. de Wit, J. Hoppe and H. Nicolai, {\it  On the Quantum Mechanics of Supermembranes},
 Nucl. Phys. {\bf B305} [FS23]
(1988) 545.

\bibitem{matrix} E. Witten, {\it Bound States Of Strings And $p$-Branes},   hep-th/9510135;
Nucl. Phys. {\bf B460} (1996) 335.

\bibitem{GSW} M.B. Green, J.H. Schwarz and E. Witten,
{\it Superstring Theory}, Cambridge U. Press, 1987.

\bibitem{b} C. Bachas, {\it D-Brane Dynamics}, hep-th/9511043;
 Phys. Lett. {\bf B374}
(1996) 37. 

\bibitem{J2} S. Chaudhuri, C. Johnson and J. Polchinski, {\it Notes on
D-Branes}, hep-th/9602052; J.  Polchinski, {\it TASI lectures on
D-branes},
hep-th/9611050.

\bibitem{romans} L. Romans, {\it Massive $N=2A$ Supergravity in Ten Dimensions},
 Phys. Lett. {\bf 169B} (1986) 374.

\bibitem{joe} J. Polchinski,
{\it Dirichlet-Branes and Ramond-Ramond Charges}, hep-th/9510017;
 Phys.Rev.Lett. {\bf 75} (1995) 4724.

\bibitem{bgpt} E. Bergshoeff, M. de Roo, M. B. Green, G. Papadopoulos and
 P. K. Townsend, {\it Duality of Type II 7-branes and 8-branes},   hep-th/9601150;
Nucl.Phys. {\bf B470} (1996) 113.

\bibitem{callan}C.G. Callan and J.A.  Harvey,
{\it Anomalies and Fermion Zero Modes on Strings and Domain Walls},
Nucl. Phys. {\bf B250} (1985) 427.


\bibitem{ghm} M.B.  Green, J.A.  Harvey and G.  Moore, {\it I-brane
Inflow and
Anomalous Couplings on D-branes}, hep-th/9605033; Class. Quant.
Grav.  {\bf 14}
(1997) 47.

\bibitem{deroo}  E.  Bergshoeff and M. De Roo, {\it D-branes and T duality}, hep-th/9603123; 
Phys. Lett. {\bf B380} (1996) 265.
 

\bibitem{seibmor} N. Seiberg and D.R. Morrison, {\it
Extremal Transitions and Five-Dimensional Supersymmetric Field Theories},  hep-th/9609070;
Nucl.Phys. {\bf B483}  (1997) 229.


\end{thebibliography}
\end{document}